

{\def\unredoffs{\hoffset-.14truein\voffset-.21truein
                     \vsize=9.5truein}
\newbox\leftpage \newdimen\fullhsize \newdimen\hstitle \newdimen\hsbody
\tolerance=1000\hfuzz=2pt
\catcode`\@=11 
\def\bigans{b }
\magnification=1100\unredoffs\baselineskip=16pt plus 2pt minus 1pt
\hsbody=\hsize\hstitle=\hsize
\def\almostshipout#1{
\count1=1 \message{[\the\count0.\the\count1]}
      \global\setbox\leftpage=#1 \global\let\l@r=R}
\newcount\yearltd\yearltd=\year\advance\yearltd by -1900

%
%

\def\draftmode{\message{ DRAFTMODE }\def\draftdate{{\rm preliminary draft:
\number\month/\number\day/\number\yearltd\ \ \hourmin}}%
\headline={\hfil\draftdate}\writelabels\baselineskip=10pt plus 2pt minus 2pt
 {\count255=\time\divide\count255 by 60 \xdef\hourmin{\number\count255}
  \multiply\count255 by-60\advance\count255 by\time
  \xdef\hourmin{\hourmin:\ifnum\count255<10 0\fi\the\count255}}}
\def\nolabels{\def\wrlabeL##1{}\def\eqlabeL##1{}\def\reflabeL##1{}}
\def\writelabels{\def\wrlabeL##1{\leavevmode\vadjust{\rlap{\smash%
{\line{{\escapechar=` \hfill\rlap{\sevenrm\hskip.03in\string##1}}}}}}}%
\def\eqlabeL##1{{\escapechar-1\rlap{\sevenrm\hskip.05in\string##1}}}%
\def\reflabeL##1{\noexpand\llap{\noexpand\sevenrm\string\string\string##1}}}
\nolabels
%
\global\newcount\secno \global\secno=0
\global\newcount\meqno \global\meqno=1
\def\newsec#1{\global\advance\secno by1\message{(\the\secno. #1)}
\global\subsecno=0\eqnres@t\noindent{\bf\the\secno. #1}
\writetoca{{\secsym} {#1}}\par\nobreak\medskip\nobreak}
\def\eqnres@t{\xdef\secsym{\the\secno.}\global\meqno=1\bigbreak\bigskip}
\def\sequentialequations{\def\eqnres@t{\bigbreak}}\xdef\secsym{}
\global\newcount\subsecno \global\subsecno=0
\def\subsec#1{\global\advance\subsecno by1\message{(\secsym\the\subsecno. #1)}
\ifnum\lastpenalty>9000\else\bigbreak\fi
\noindent{\it\secsym\the\subsecno. #1}\writetoca{\string\quad
{\secsym\the\subsecno.} {#1}}\par\nobreak\medskip\nobreak}
\def\appendix#1#2{\global\meqno=1\global\subsecno=0\xdef\secsym{\hbox{#1.}}
\bigbreak\bigskip\noindent{\bf Appendix #1. #2}\message{(#1. #2)}
\writetoca{Appendix {#1.} {#2}}\par\nobreak\medskip\nobreak}
%
%
\def\eqnn#1{\xdef #1{(\secsym\the\meqno)}\writedef{#1\leftbracket#1}%
\global\advance\meqno by1\wrlabeL#1}
\def\eqna#1{\xdef #1##1{\hbox{$(\secsym\the\meqno##1)$}}
\writedef{#1\numbersign1\leftbracket#1{\numbersign1}}%
\global\advance\meqno by1\wrlabeL{#1$\{\}$}}
\def\eqn#1#2{\xdef #1{(\secsym\the\meqno)}\writedef{#1\leftbracket#1}%
\global\advance\meqno by1$$#2\eqno#1\eqlabeL#1$$}
%
\newskip\footskip\footskip14pt plus 1pt minus 1pt 
\def\footnotefont{\ninepoint}\def\f@t#1{\footnotefont #1\@foot}
\def\f@@t{\baselineskip\footskip\bgroup\footnotefont\aftergroup\@foot\let\next}
\setbox\strutbox=\hbox{\vrule height9.5pt depth4.5pt width0pt}
\global\newcount\ftno \global\ftno=0
\def\foot{\global\advance\ftno by1\nobreak\footnote{$^{\the\ftno}$}}
%
\newwrite\ftfile
\def\footend{\def\foot{\global\advance\ftno by1\chardef\wfile=\ftfile
$^{\the\ftno}$\ifnum\ftno=1\immediate\openout\ftfile=foots.tmp\fi%
\immediate\write\ftfile{\noexpand\smallskip%
\noexpand\item{f\the\ftno:\ }\pctsign}\findarg}%
\def\footatend{\vfill\eject\immediate\closeout\ftfile{\parindent=20pt
\centerline{\bf Footnotes}\nobreak\bigskip\input foots.tmp }}}
\def\footatend{}
%
%
\global\newcount\refno \global\refno=1
\newwrite\rfile
\def\ref{\nobreak[\the\refno]\nref}
\def\nref#1{\xdef#1{[\the\refno]}\writedef{#1\leftbracket#1}%
\ifnum\refno=1\immediate\openout\rfile=refs.tmp\fi
\global\advance\refno by1\chardef\wfile=\rfile\immediate
\write\rfile{\noexpand\item{#1\ }\reflabeL{#1\hskip.31in}\pctsign}\findarg}
\def\findarg#1#{\begingroup\obeylines\newlinechar=`\^^M\pass@rg}
{\obeylines\gdef\pass@rg#1{\writ@line\relax #1^^M\hbox{}^^M}%
\gdef\writ@line#1^^M{\expandafter\toks0\expandafter{\striprel@x #1}%
\edef\next{\the\toks0}\ifx\next\em@rk\let\next=\endgroup\else\ifx\next\empty%
\else\immediate\write\wfile{\the\toks0}\fi\let\next=\writ@line\fi\next\relax}}
\def\striprel@x#1{} \def\em@rk{\hbox{}}
\def\lref{\begingroup\obeylines\lr@f}
\def\lr@f#1#2{\gdef#1{\ref#1{#2}}\endgroup\unskip}
\def\semi{;\hfil\break}
\def\addref#1{\immediate\write\rfile{\noexpand\item{}#1}} 
\def\startrefs#1{\immediate\openout\rfile=refs.tmp\refno=#1}
\def\xref{\expandafter\xr@f}\def\xr@f[#1]{#1}
\def\refs#1{\count255=1\nobreak[\r@fs #1{\hbox{}}]}
\def\r@fs#1{\ifx\und@fined#1\message{reflabel \string#1 is undefined.}%
\nref#1{need to supply reference \string#1.}\fi%
\vphantom{\hphantom{#1}}\edef\next{#1}\ifx\next\em@rk\def\next{}%
\else\ifx\next#1\ifodd\count255\relax\xref#1\count255=0\fi%
\else#1\count255=1\fi\let\next=\r@fs\fi\next}
%

%
\newwrite\ffile\global\newcount\figno \global\figno=1
\def\fig{fig.~\the\figno\nfig}
\def\nfig#1{\xdef#1{fig.~\the\figno}%
\writedef{#1\leftbracket fig.\noexpand~\the\figno}%
\ifnum\figno=1\immediate\openout\ffile=figs.tmp\fi\chardef\wfile=\ffile%
\immediate\write\ffile{\noexpand\medskip\noexpand\item{Fig.\ \the\figno. }
\reflabeL{#1\hskip.55in}\pctsign}\global\advance\figno by1\findarg}
\def\xfig{\expandafter\xf@g}\def\xf@g fig.\penalty\@M\ {}
\def\figs#1{figs.~\f@gs #1{\hbox{}}}
\def\f@gs#1{\edef\next{#1}\ifx\next\em@rk\def\next{}\else
\ifx\next#1\xfig #1\else#1\fi\let\next=\f@gs\fi\next}
\newwrite\lfile
{\escapechar-1\xdef\pctsign{\string\%}\xdef\leftbracket{\string\{}
\xdef\rightbracket{\string\}}\xdef\numbersign{\string\#}}

\def\writestop{\def\writestoppt{\immediate\write\lfile{\string\pageno%
\the\pageno\string\startrefs\leftbracket\the\refno\rightbracket%
\string\def\string\secsym\leftbracket\secsym\rightbracket%
\string\secno\the\secno\string\meqno\the\meqno}\immediate\closeout\lfile}}
\def\writestoppt{}\def\writedef#1{}
\def\seclab#1{\xdef #1{\the\secno}\writedef{#1\leftbracket#1}\wrlabeL{#1=#1}}
\def\subseclab#1{\xdef #1{\secsym\the\subsecno}%
\writedef{#1\leftbracket#1}\wrlabeL{#1=#1}}
\newwrite\tfile \def\writetoca#1{}
\def\leaderfill{\leaders\hbox to 1em{\hss.\hss}\hfill}
\def\writetoc{\immediate\openout\tfile=toc.tmp
   \def\writetoca##1{{\edef\next{\write\tfile{\noindent ##1
   \string\leaderfill {\noexpand\number\pageno} \par}}\next}}}

%
\catcode`\@=12 
%
\edef\tfontsize{scaled\magstep3}
 \tfontsize  \tfontsize
 \tfontsize \font\titlei=cmmi10 \tfontsize
\font\titleis=cmmi7 \tfontsize \font\titleiss=cmmi5 \tfontsize
\font\titlesy=cmsy10 \tfontsize \font\titlesys=cmsy7 \tfontsize
\font\titlesyss=cmsy5 \tfontsize  \tfontsize
\skewchar\titlei='177 \skewchar\titleis='177 \skewchar\titleiss='177
\skewchar\titlesy='60 \skewchar\titlesys='60 \skewchar\titlesyss='60
\font\ninerm=cmr9 \font\sixrm=cmr6 \font\ninei=cmmi9 \font\sixi=cmmi6
\font\ninesy=cmsy9 \font\sixsy=cmsy6 \font\ninebf=cmbx9
\font\nineit=cmti9 \font\ninesl=cmsl9 \skewchar\ninei='177
\skewchar\sixi='177 \skewchar\ninesy='60 \skewchar\sixsy='60
\def\ninepoint{\def\rm{\fam0\ninerm}
\textfont0=\ninerm \scriptfont0=\sixrm \scriptscriptfont0=\fiverm
\textfont1=\ninei \scriptfont1=\sixi \scriptscriptfont1=\fivei
\textfont2=\ninesy \scriptfont2=\sixsy \scriptscriptfont2=\fivesy
\textfont\itfam=\ninei \def\it{\fam\itfam\nineit}\def\sl{\fam\slfam\ninesl}%
\textfont\bffam=\ninebf \def\bf{\fam\bffam\ninebf}\rm}
%
%
\def\noblackbox{\overfullrule=0pt}
\def\inv{^{\raise.15ex\hbox{${\scriptscriptstyle -}$}\kern-.05em 1}}

\def\Dsl{\,\raise.15ex\hbox{/}\mkern-13.5mu D} 
\def\dsl{\raise.15ex\hbox{/}\kern-.57em\partial}

\def\lspace{\ifx\answ\bigans{}\else\qquad\fi}
\def\lbspace{\ifx\answ\bigans{}\else\hskip-.2in\fi} 
\def\boxeqn#1{\vcenter{\vbox{\hrule\hbox{\vrule\kern3pt\vbox{\kern3pt
	\hbox{${\displaystyle #1}$}\kern3pt}\kern3pt\vrule}\hrule}}}
\def\mbox#1#2{\vcenter{\hrule \hbox{\vrule height#2in
		\kern#1in \vrule} \hrule}}  
\def\tilde{\widetilde} \def\bar{\overline} \def\hat{\widehat}


\noblackbox

\pageno=0\nopagenumbers\tolerance=10000\hfuzz=5pt
\line{\hfill CERN-TH.7234/94}
\line{\hfill RAL-94-039}
\vskip 36pt
\centerline{\bf THE DECOUPLING THEOREM IN}
\centerline{\bf EFFECTIVE SCALAR FIELD THEORY}
\vskip 36pt\centerline{Richard~D.~Ball$^{a}$\footnote*{On leave of
absence from a Royal Society University Research Fellowship.}
and Robert~S.~Thorne$^{b}$}
\vskip 12pt
\centerline{\it Theory Division, CERN,}
\centerline{\it CH-1211 Gen\`eve 23, Switzerland.$^{a}$}
\vskip 10pt
\centerline{\it and }
\vskip 10pt
\centerline{\it Rutherford Appleton Laboratory,}
\centerline{\it Chilton, Didcot, Oxon., OX11 0QX, U.K.~$^{b}$}
\vskip 1.in
{\narrower\baselineskip 10pt
\centerline{\bf Abstract}
\medskip
We consider decoupling in the context of an effective quantum
field theory of two scalar fields with well separated mass scales and
a $Z_2\times Z_2$ symmetry. We first prove, using Wilson's exact
renormalization group equation, that the theory is renormalizable, in
the same way that we showed in a previous paper that theories with a
single mass scale were renormalizable. We then state and prove a
decoupling theorem: at scales below the mass of the heavy particle
the full theory may be approximated arbitrarily closely by an
effective theory of the light particle alone, with naturalness scale
the heavy particle mass. We also compare our formulation of effective
field theory with the more conventional local formulation.
}

\vskip 0.8in
\line{CERN-TH.7234/94\hfill}
\line{RAL-94-039\hfill}
\line{April 1994\hfill}
\vfill\eject
\footline={\hss\tenrm\folio\hss}


\def\frac#1#2{{{#1}\over {#2}}}
\def\half{\hbox{${1\over 2}$}}

\def\smallfrac#1#2{\hbox{${{#1}\over {#2}}$}}

\def\blackbox{\vrule height7pt width5pt depth2pt}

\catcode`@=11 
\def\slash#1{\mathord{\mathpalette\c@ncel#1}}
 \def\c@ncel#1#2{\ooalign{$\hfil#1\mkern1mu/\hfil$\crcr$#1#2$}}
\def\lsim{\mathrel{\mathpalette\@versim<}}
\def\gsim{\mathrel{\mathpalette\@versim>}}
 \def\@versim#1#2{\lower0.2ex\vbox{\baselineskip\z@skip\lineskip\z@skip
       \lineskiplimit\z@\ialign{$\m@th#1\hfil##$\crcr#2\crcr\sim\crcr}}}
\catcode`@=12 

\def\PR{{\it Phys.~Rev.~}}
\def\PRL{{\it Phys.~Rev.~Lett.~}}
\def\NP{{\it Nucl.~Phys.~}}
\def\PL{{\it Phys.~Lett.~}}
\def\PRep{{\it Phys.~Rep.~}}
\def\AP{{\it Ann.~Phys.~}}
\def\CMP{{\it Comm.~Math.~Phys.~}}
\def\JMP{{\it Jour.~Math.~Phys.~}}

\def\vol#1{{\bf #1}}
\def\vyp#1#2#3{\vol{#1} (#2) #3}


\nref\REFT{R.~D.~Ball and R.~S.~Thorne, OUTP-93-23P, CERN-TH.7067/93,
to be published in {\it Ann.~Phys.}}
\nref\rWil{K.~G.~Wilson and J.~G.~Kogut, \PRep\vyp{12C}{1974}{75}.}
\nref\rPol{J. Polchinski, \NP\vyp{B231}{1984}{269}.}

In a previous paper \REFT\ we considered an effective
scalar quantum field theory containing
a single propagating scalar particle with a $Z_2$ global symmetry.
We constructed a regulated classical action for
this theory which is expandable as a convergent infinite series of
local terms, and is characterized by some such naturalness scale,
$\Lambda_0$, at which all the coupling constants in the action are
of order one when expressed in units of $\Lambda_0$. The particle mass
$m$ is assumed to be fine--tuned so that $m\ll\Lambda_0$, so that we
can consider scattering amplitudes at energy scales $E\ll\Lambda_0$.
We were then able to show that not only is the S--matrix of the
theory well--defined, unitary and causal at all scales, but that the
theory is perturbatively renormalizable, in the sense that
these low--energy scattering amplitudes can be expressed with great
accuracy in terms of only a finite number of physically relevant
parameters. Technically this separation of scales was
achieved through a systematic
exploitation of Wilson's exact renormalization group equation
\refs{\rWil,\rPol}.
All the results may be readily generalized to theories with many scalar
or spin--half particles, possibly with linearly realized global
symmetries, provided all the particles have masses $m_i$ of the same order
of magnitude.

Now as the energy scale $E$ at which the physics of the
theory is being probed is increased, we were able to show further
that the degree of predictivity of the theory can be maintained by
including more and more measurable
parameters (by fixing more and more low energy renormalization
conditions on couplings). We could also derive Weinberg--style
bounds on the growth with $E$ of Euclidean Green's functions
\ref\Infrared{R.~D.~Ball and R.~S.~Thorne, CERN-TH.????/94,
RAL-94-039.}. For energies of order $\Lambda_0$ we need formally an
infinite number of
parameters and have no real predictive power beyond that offered by
S--matrix theory; indeed it was possible to show that the effective
field theory and S--matrix theory are formally equivalent.

As physicists we are sufficiently optimistic to hope that the
naturalness scale $\Lambda_0$ will be the scale at which we will find
some sort of new physics which will serve to substantially restrict
the infinite number of possible theories to some more manageable
finite subset, so as to regain a reasonably predictive theory over a
new range of energy scales. However, the effective theory as
considered so far is phenomenologically neutral, in the sense that
by construction it is completely
independent of the form this new physics may take, provided only that
it be consistent with the basic principles of special relativity and
quantum mechanics. Indeed, this is why there are so many possible
effective theories of even a single scalar particle. So at $\Lambda_0$
it is desirable to put forward additional assumptions. Here we will
consider what is probably the simplest such assumption: that a new
scalar particle of mass $M$ of the order of $\Lambda_0$ is discovered,
and that the naturalness scale of the new theory can then be pushed, by further
fine tuning, to even higher scales. We are thus led to consider the
following simple paradigm; a theory
containing two scalar particles with significantly different masses $m\ll
M\ll \Lambda_0$, in which we can then determine the dependence of the
physics at various scales depends on the ratio of the
masses of the particles. In particular, we will be able to see exactly
how the heavy particle `decouples' from the low energy
effective theory containing only the light particle.

Decoupling in local quantum field theories was first investigated in
the early 1970's, with the development of a decoupling theorem
\ref\rdi{K.~Symanzik, \CMP \vyp{18}{1970}{7}\semi
S.~Weinberg, \PR\vyp{D8}{1973}{605}\semi
T.~Appelquist and J.~Carazzone, \PR \vyp{D11}{1975}{2856}.}
which stated that if we have a local quantum field theory in which some
particles have masses $M$ very large compared to the others then
Green's functions for light particle processes at energies $E\ll M$
are the same as those in the local theory
obtained by simply omitting the heavy particles, up
to corrections of inverse powers of the heavy mass. The only effect of
the heavy particles is thus to renormalize the masses and relevant
couplings in the light particle theory.
Rigorous proofs \ref\ja{J.~Ambj{\o}rn \CMP \vyp{67}{1979}{109}
\semi E.~B.~Manoukian, \JMP\vyp{22}{1981}{572,2258}
\semi T.~Schimert and C.~C.~Chiang, \PR\vyp{D29}{1984}{241}.}
of this decoupling theorem using the BPHZ subtraction
formalism took some time to develop, however,
essentially because in this formalism one attempts to deal
with all the different scales, including of course the cut--off $\Lambda_0$,
at the same time. If the light particles are massless, there are
further complications, and extensions of the decoupling
theorem to include the leading order $E^2/M^2$
corrections \ref\lky{C.~Lee, \NP\vyp{B161}{1979}{171}\semi
Y.~Kazama and Y.~P.~Yao, \PR\vyp{D21}{1980}{1116}.} become
increasingly awkward to deal with satisfactorily.

We will find here that on the contrary, since exact
renormalization group techniques are
very good at separating the physics at different scales, it is relatively
straightforward to use them to prove
an extended decoupling theorem which can treat corrections to the
light particle effective theory of arbitrarily high orders in
$E^2/M^2$.  In effect the theorem reduces to a statement of complex
analysis: the light particle vertex functions of the full theory,
which have Taylor expansions with radius of convergence $M$, may be
approximated to arbitrary accuracy within their circle of convergence
by the regular vertex functions of the light theory.

\nref\eftdec{
H.~Georgi, H.~R.~Quinn and S.~Weinberg, \PRL\vyp{33}{1974}{451}\semi
E.~Witten, \NP\vyp{B104}{1976}{445}\semi
S.~Weinberg, \PL\vyp{91B}{1980}{51}\semi
B.~A.~Ovrut and H.~J.~Schnitzer,
\PR\vyp{D21}{1980}{3369};\ \vyp{D22}{1980}{2518};\
\vyp{D24}{1981}{1695};\ \NP\vyp{B184}{1981}{109}\semi
H.~Georgi, \NP\vyp{B232}{1984}{61}
;\ \vyp{B363}{1991}{301};\ {\it (Proc. Suppl.)~}\vyp{29B,C}{1992}{1}.}
This relatively straightforward interpretation of decoupling is also
due in part to our particular formulation of effective field theory in
\REFT. In the more conventional approach to effective theories
which was in fact motivated in part by precisely the sort of
decoupling situation
discussed here\refs{\eftdec}, the effective theory of the light
particle is always considered only in the strictly local limit in
which the cut--off
is removed. This makes it difficult to entertain the possibility of an
effective theory which is not tied in some way to an underlying local
`fundamental' theory. For us the light particle theory is simply an
effective theory with naturalness scale of order $M$, to be (perhaps)
superseded at scales of order $M$ by another effective theory with
naturalness scale $\Lambda_0\gg M$, to which it is matched.

Once we have defined precisely the full theory we are working with,
with the three well--separated scales $m\ll M\ll\Lambda_0$, our
first step will be to prove the boundedness and convergence of this
theory (\S 2). Once we have done this it will be relatively easy
to prove in \S 3 the conventional form of the decoupling theorem. In \S 4
we will then show how we can systematically improve this decoupling
theorem to make the predictions of the low energy theory arbitrarily
accurate at a given low energy scale, and finally in \S 5 we discuss
the relation between our formulation of effective theories and the
more conventional one.

\newsec{Defining the Theory}

We will prove the decoupling theorem for the simplest case, i.e. a
single light scalar particle described by the field $\phi$, and
a single heavy scalar particle
described by the field $\Phi$, where the theory is invariant under
a $Z_2\times Z_2$ global symmetry, under which the fields transform as
$\phi \rightarrow -\phi$,
$\Phi \rightarrow -\Phi$.

We must first define the theory containing
these two particles. As explained in \REFT, the effective quantum
field theory is defined by its field content, global symmetries, the
form of the propagators, and a full set of boundary conditions on the
renormalization group flow for the interaction part of the Lagrangian.
As in \REFT\ we separate the Euclidean classical action into a quadratic
part and an interaction:
\eqn\eclassact
{S[\phi,\Phi;\Lambda_0] = \half\big(\phi,P_\Lambda\inv\phi\big)+
\half\big(\Phi,\tilde P_\Lambda\inv\Phi\big)
+ S_{\rm int}[\phi,\Phi;\Lambda_0].}
The propagators are defined by
\eqn\edi{\eqalign{P_\Lambda (p)
&={K((p^2 + m^2)/\Lambda^2) \over (p^2 +m^2)}
={K_\Lambda (p) \over (p^2 +m^2)},\cr
\tilde P_\Lambda (p)&={\tilde K((p^2 + M^2)/\Lambda^2) \over (p^2 +M^2),}
= {\tilde K_\Lambda (p) \over (p^2 +M^2)},\cr}}
where $K(x), \tilde K(x)$ are regular regulating functions with the same
properties as those used in \REFT.\foot{In
practice it would be simpler to take $K=\tilde K$, but this is not
necessary.}
The interaction part of the action is defined to be power series in
the fields $\phi$ and $\Phi$; because of the assumed $Z_2\times Z_2$ global
symmetry only even powers appear. For perturbative purposes it is also
assumed to be a formal power series in three small expansion
parameters, $g_m$, $g_M$, $g_{mM}$, beginning at first order in at
least one of these
($g_m$, $g_M$ and $g_{mM}$ will normally be related to the coupling
constants of the theory). So
\eqn\eint{\eqalign{
S_{\rm int}[\phi,\Phi;\Lambda] \equiv \sum^{\infty}_{m,n=0}
&\sum^{\infty}_{r,s,t=0}{{g_m^r g_M^s g_{mM}^t}\over{(2(m+n))!}}
\int {{d^4p_1\cdots d^4p_{2(m+n)}}\over{(2\pi)^{4(2(m+n) - 1)}}}\cr
&V^{r,s,t}_{2m,2n}(p_1,\ldots ,p_{2(m+n)};\Lambda)
\delta^4\big(\hbox{$\sum^{2(m+n)}_{i=1}p_i$}\big)
\phi_{p_1}\cdots\phi_{p_{2m}}\Phi_{p_{2m+1}}\cdots\Phi_{p_{2(m+n)}},\cr}}
where $V^{r,s,t}_{2m,2n}(p_1,\ldots ,p_{2(m+n)};\Lambda)
\equiv V^{r,s,t}_{2m,2n}(p_i;\Lambda)$ is the value, at order $r$ in
$g_m$, order $s$ in $g_M$ and order $t$ in $g_{mM}$, of the vertex in
the effective action defined at $\Lambda$ with $2m$  $\phi$-legs and
$2n$ $\Phi$-legs; $V^{r,s,t}_{2m,2n}(p_i;\Lambda)\equiv 0$ if
$r=s=t=0$ or if $m=n=0$.

We define a complete set of boundary
conditions on the renormalization group flow  by setting all of the irrelevant
couplings at $\Lambda_0$, the naturalness scale for the full theory,
equal to zero (this assumption will be relaxed in \S 2.2 below),
\eqn\exxxv{\partial^j_p V^{r,s,t}_{2m,2n}(\Lambda_0) = 0
\qquad 2(m+n) + j > 4,}
and choosing the relevant couplings at $\Lambda=0$ to be equal to
$\Lambda_0$-independent constants (at each
order in perturbation theory):
\eqn\eIxv
{\eqalign{\lim_{\Lambda\to 0}V^{r,s,t}_{2,0}(P_0,-P_0;\Lambda) &=
\Lambda_m^2\hat\lambda^{r,s,t}_{1},\cr
\lim_{\Lambda\to 0}V^{r,s,t}_{0,2}(\tilde{P_0},-\tilde{P_0};\Lambda) &=
\Lambda_M^2\hat\lambda^{r,s,t}_{4},\cr}\qquad
\eqalign{\lim_{\Lambda\to 0}\bigl[
\partial_{p_{\mu}}\partial_{p_{\nu}}V^{r,s,t}_{2,0}(p,-p;\Lambda)
\vert_{p=P_0}\bigr]_{\delta_{\mu\nu}} &=\hat\lambda^{r,s,t}_{2},\cr
\lim_{\Lambda\to 0}\bigl[
\partial_{p_{\mu}}\partial_{p_{\nu}}V^{r,s,t}_{0,2}(p,-p;\Lambda)
\vert_{p=\tilde{P_0}}\bigr]_{\delta_{\mu\nu}} &=\hat\lambda^{r,s,t}_{5},\cr}}
$${V^{r,s,t}_{4,0}(P_1,P_2,P_3,P_4;0) =
\hat\lambda^{r,s,t}_{3}\! ,\quad
V^{r,s,t}_{0,4}(\tilde{P}_1,\tilde{P}_2,\tilde{P}_3,\tilde{P}_4;0) =
\hat\lambda^{r,s,t}_{6}\! ,\quad
V^{r,s,t}_{2,2}(P_5,P_6,\tilde{P}_5,\tilde{P}_6;0) =
\hat\lambda^{r,s,t}_{7}\! ,}$$
where $\sum_1^4 P_i=\sum_1^4 \tilde{P}_i=\sum_{5,6}
(P_i+\tilde{P}_i)=0$. By convention the series expansions for
those couplings corresponding to vertices  with $ m=0$ are such that
$r=0, t=0$, those for $n=0$ have $s=0, t=0$  and for both $m\neq 0$
and $n\neq 0$, $r=s=0$. We choose the external momenta $P_i$ with
magnitudes of order $\Lambda_m$, where $\Lambda_m \sim m$, for the
legs of vertices corresponding to $\phi$ fields,
and the external momenta $\tilde{P}_i$ with magnitudes of order $\Lambda_M$,
where $\Lambda_M \sim M$,
for the legs of vertices corresponding to $\Phi$ fields; the
only vertex where we have a mixture of magnitudes of momenta is then
the vertex with two $\phi$ legs and two $\Phi$ legs, where we have two
momenta $\sim \Lambda_m$ and two momenta $\sim \Lambda_M$. In this way
the renormalization conditions \eIxv\ may be readily continued
on--shell, where they would correspond to physically observable masses
and couplings. Setting all the renormalization conditions for momenta
of the same order of magnitude would be artificial, in the sense that
the relation between the renormalized `masses' and `couplings'
$\lambda_i\equiv\sum_{r,s,t}\hat\lambda^{r,s,t}_{i}$ and the
physical parameters could  (and in general will) involve
large renormalization factors, depending on the ratio
$\Lambda_M/\Lambda_m\sim M/m$.

The quantum theory of these two massive particles is then defined as
in \REFT\ by reducing $\Lambda$ from the naturalness scale $\Lambda_0$
through the two mass scales $\Lambda_M$ and $\Lambda_m$ down to zero,
while keeping the connected amputated Green's functions $\tilde
G^c_{2m,2n}$; the interaction $S_{\rm int}[\phi,\Phi;\Lambda]$ must
the satisfy the exact renormalization group equation
\eqn\eiix
{{\partial S_{\rm int} \over \partial \Lambda}
= \half\int {d^4 p\over (2 \pi )^4}
\biggl\{ {\partial P_\Lambda \over\partial\Lambda}
      \biggl[{\delta S_{\rm int}\over\delta\phi_p}
             {\delta S_{\rm int}\over\delta\phi_{-p}}
           - {\delta^2 S_{\rm int}\over\delta\phi_p\delta\phi_{-p}}\biggr]
       + {\partial \tilde P_\Lambda \over\partial\Lambda}
      \biggl[{\delta S_{\rm int}\over\delta\Phi_p}
             {\delta S_{\rm int}\over\delta\Phi_{-p}}
           - {\delta^2 S_{\rm int}\over\delta\Phi_p\delta\Phi_{-p}}\biggr]
\biggr\}.}
The connected amputated Green's functions may then be read off directly
from $S_{\rm int}[\phi,\Phi;0]$:
\eqn\eacgf{\tilde{G}^{c}_{2m,2n}(p_1,\ldots ,p_{2(m+n)})\equiv
\prod_{i=1}^{2m}\Big(-\frac{\delta}{\delta \phi_{p_i}}\Big)
\prod_{j=1}^{2n}\Big(-\frac{\delta}{\delta \Phi_{p_j}}\Big)
S_{\rm int}[\phi,\Phi;0]\Big\vert_{\phi=\Phi=0}.}

\newsec{Renormalizability.}

We now consider the theory described in the previous section when
all three scales are well separated: $\Lambda_m \ll \Lambda_M
\ll\Lambda_0$. We wish to derive bounds on Green's functions analogous
to those proved in \REFT\ for the theory with only a single mass scale
(i.e. with $\Lambda_m\sim\Lambda_M\ll\Lambda_0$), and then show
further that the Green's functions are convergent (independent of
$\Lambda_0$ up to power suppressed corrections) and universal
(independent of the boundary conditions at $\Lambda_0$ on irrelevant
vertices, again up to power suppressed corrections). We will then
have proven that the theory is `renormalizable', and can proceed to an
investigation of decoupling.

\subsec{Boundedness.}

In order to derive appropriate bounds, we use similar methods to those
in used in \REFT\ for lemmas 1--3; we first derive bounds on the vertices
and their derivatives using the flow equation \eiix, and then proceed
by induction in the order of the vertices and the number of their legs.
The norm on the vertices is now defined as
\eqn\eNi{\Vert V_{2m,2n}\Vert_{\Lambda_1,\Lambda_2} \equiv
\max_{\{p_1,\ldots ,p_{2(m+n)}\}}\biggl[\prod_{i=1}^{2m}
[K_{\Lambda_1}(p_i)]^{1/4}\prod_{j=1}^{2n}[\tilde K_{\Lambda_2}(p_j)]^{1/4}
\vert V(p_1,...,p_{2(m+n)};\Lambda)\vert\biggr].}
If a vertex does not have legs corresponding to one of the fields
(i.e. if either $m$ or $n$ is zero) we will omit the superscript
referring to the damping factors attached to these legs.

The flow equation for the vertices is obtained by substituting \eint\
into \eiix\ and we can bound the $j_{th}$ momentum derivative of the
left hand side to give
\eqn\ediv{\eqalign{\Biggl \Vert  &{\partial \over \partial\Lambda} \biggl(
\partial^j_p V^{r,s,t}_{2m,2n}(\Lambda) \biggr) \Biggr\Vert_{\Lambda,\Lambda}
\leq \Lambda \Vert  \partial_p^j V^{r,s,t}_{2m+2,2n}(\Lambda)
\Vert_{\Lambda,\Lambda}+\Lambda\Vert
\partial_p^j V^{r,s,t}_{2m,2n+2}(\Lambda)
\Vert_{\Lambda,\Lambda}\cr
&+ \sum^{r-1}_{r'=1} \sum^{s-1}_{s'=1} \sum
^{t-1}_{t'=1} \sum_{j_i;j_1 +j_2 +j_3=j}
\Big[\sum^m_{l=1} \sum^n_{k=0}  \Lambda^{-3-j_1}\Vert
\partial_p^{j_2}V^{r',s',t'}_{2l,2k}(\Lambda)
\Vert_{\Lambda,\Lambda} \cdot \Vert\partial^{j_3}_p
V^{r-r',s-s',t-t'}_{2m + 2 -2l,2n-2k}(\Lambda)
\Vert_{\Lambda,\Lambda} \cr
&\qquad\qquad +\sum^m_{l=0}\sum^n_{k=1}
\Lambda^{-3-j_1}\Vert
\partial_p^{j_2}V^{r',s',t'}_{2l,2k}(\Lambda)
\Vert_{\Lambda,\Lambda} \cdot \Vert
\partial_p^{j_3}V^{r-r',s-s',t-t'}_{2m -2l,2n-2k+2}(\Lambda)
\Vert_{\Lambda,\Lambda}\Big],\cr}}
for $\Lambda\in [\Lambda_M,\Lambda_0]$. For $\Lambda \in
[0,\Lambda_M]$ we find instead
\eqn\edivi{\eqalign{\Biggl \Vert  &{\partial \over \partial\Lambda} \biggl(
\partial^j_p V^{r,s,t}_{2m,2n}(\Lambda) \biggr)
\Biggr\Vert_{\Lambda',\Lambda_M}
\leq \Lambda \Vert  \partial_p^j V^{r,s,t}_{2m+2,2n}(\Lambda)
\Vert_{\Lambda',\Lambda_M}+\Lambda\Vert
\partial_p^j V^{r,s,t}_{2m,2n+2}(\Lambda)
\Vert_{\Lambda',\Lambda_M}\cr
&+ \sum^{r-1}_{r'=1} \sum^{s-1}_{s'=1} \sum
^{t-1}_{t'=1} \sum_{j_i;j_1 +j_2 +j_3=j}
\Big[\sum^m_{l=1} \sum^n_{k=0}  \Lambda^{-3-j_1}\Vert
\partial_p^{j_2}V^{r',s',t'}_{2l,2k}(\Lambda)
\Vert_{\Lambda',\Lambda_M} \cdot \Vert\partial^{j_3}_p
V^{r-r',s-s',t-t'}_{2m + 2 -2l,2n-2k}(\Lambda)
\Vert_{\Lambda',\Lambda_M} \cr
&\qquad +\sum^m_{l=0}\sum^n_{k=1}
\Lambda^{-3-j_1}K^{1/2}(\hbox{${M^2\over \Lambda^2}$})\Vert
\partial_p^{j_2}V^{r',s',t'}_{2l,2k}(\Lambda)
\Vert_{\Lambda',\Lambda_M} \cdot \Vert
\partial_p^{j_3}V^{r-r',s-s',t-t'}_{2m -2l,2n-2k+2}(\Lambda)
\Vert_{\Lambda',\Lambda_M}\Big],\cr}}
where $\Lambda'={\rm max}(\Lambda,\Lambda_m)$: for $\Lambda\in
[\Lambda_m,\Lambda_M]$ we take $\Lambda'=\Lambda$, while for $\Lambda\in
[0,\Lambda_m]$ we take $\Lambda'=\Lambda_m$. We have explicitly
retained the factor of
$K^{1/2}(M^2/\Lambda^2)$ arising from the $\Lambda$--derivative of the
propagator in the last term since it will turn out to be be important
for part of the argument.

We may then prove the following bounds:\foot{To avoid confusion we
number the lemmas in \REFT, \Infrared\ and the present
paper consecutively.}
\medskip
\noindent {\it Lemma 8:}\nobreak

\noindent i) For $\Lambda \in [\Lambda_M, \Lambda_0]$, and
for all $m$, $n$, $j$, $r$, $s$ and $t$,
\eqn\edv{ \Vert \partial^j_p V^{r,s,t}_{2m,2n}(\Lambda)\Vert_{\Lambda,\Lambda}
\leq
\Lambda^{4-2m-2n-j}\Biggl( P\log \biggl({\Lambda \over \Lambda_m}\biggr) +
{\Lambda \over \Lambda_0} P \log \biggl( {\Lambda_0 \over
\Lambda_m}\biggr)\Biggr).}
\noindent ii) For $\Lambda \in [\Lambda_m, \Lambda_M]$, and for all
$m$, $j$, $r$, $s$ and $t$, except $m=0$, $n=1$ and $j=0,1$,
\eqn\edvi{ \Vert \partial^j_p
V^{r,s,t}_{2m,2n}(\Lambda)\Vert_{\Lambda,\Lambda_M} \leq\cases{
\Lambda^{4-2m-2n-j} P\log \biggl({\Lambda_M \over
\Lambda_m}\biggr),&if $n>0$;\cr
\Lambda^{4-2m-j} P\log \biggl({\Lambda \over \Lambda_m}\biggr),
&if $n=0$.\cr}}
\noindent iii) For $\Lambda \in [0, \Lambda_m]$, and for all
$m$, $j$, $r$, $s$ and $t$, except for $m=0$, $n=1$, and $j=0,1$,
\eqn\eddvi{ \Vert \partial^j_p
V^{r,s,t}_{2m,2n}(\Lambda)\Vert_{\Lambda_m,\Lambda_M} \leq\cases{
\Lambda_m^{4-2m-2n-j} P\log \biggl({\Lambda_M \over \Lambda_m}\biggr),
&if $n>0$;\cr
\Lambda_m^{4-2m-j},&if $n=0$.\cr}}
\noindent iv) For the special case $m=0$, $n=1$, $j=0,1$,
\eqn\edc{ \Vert \partial^j_p V^{r,s,t}_{0,2}(\Lambda) \Vert
_{\Lambda_M}\leq \Lambda_M^{2-j} P \log\biggl( {\Lambda_M
\over \Lambda_m}\biggr),}
for all $\Lambda \in [0,\Lambda_M]$.
\foot{The necessity for this exceptional case is easy to understand
if we remember that $V^{r,s,t}_{0,2}(\Lambda)$ are the
corrections to the mass term of the effective Lagrangian for the heavy field.
As discussed in \S 3.2 of \REFT, we would naively expect these corrections
to be of order $\Lambda_0^2$ for all $\Lambda$, but the fine-tuned
renormalization condition \eIxv\ which specifies
the mass corrections to be of order $\Lambda_M^2$ at each order in
the two point Green's function, forces them to be of order
$\Lambda^2$ as we flow down from $\Lambda_0$ to
$\Lambda_M$. However as we flow down further the
corrections now behave as we would actually expect from the naive dimensional
argument, and are thus roughly the same for all $\Lambda \leq
\Lambda_M$. This is in contrast with the mass corrections on the light
particle, which are forced to be `unnaturally' small by their
renormalization condition, so that $V^{r,s,t}_{2,0}(\Lambda)$
satisfies ii) and iii).}
\medskip
The lemma is proven using essentially the same inductive method as we
used for lemma 1 in \REFT; the fact that we
have more than one expansion parameter causes no
problems, since we may simply perform the inductive arguments
successively in the parameters. We thus assume that the lemma
is true at order $r-1,s-1,t-1$ in the
expansion parameters and order $m+1,n+1$ in the number of legs, and
then investigate whether it is true at order $r,s,t$ and $m,n$; the
vanishing of the vertices for large enough $m$ and
$n$ for any given order in $r,s,t$, and the fact that the bounds
are trivially
true when $r=s=t=0$ then completes the induction.
Complications arise however because we had to set the
renormalization conditions \eIxv\ at momenta consistent with two
different scales, $\Lambda_m$ and $\Lambda_M$.
This causes no real difficulties if we are careful about the order in
which we bound the vertices. In fact each
inductive step now proceeds in six distinct steps: a) we prove i) for the
irrelevant vertices; b) ii) for the irrelevant vertices; c) iii) for
the irrelevant vertices; d) iii) for relevant vertices involving only
light fields, and ii), iii) and iv) those involving heavy fields;
e) ii) for the relevant vertices involving only light fields; and finally
f) i) for the relevant vertices.

\medskip

a) We first consider irrelevant vertices for $\Lambda \in
[\Lambda_M,\Lambda_0]$. The flow equations
are perfectly consistent with i), so integrating from $\Lambda
\geq \Lambda_M$ up to
$\Lambda_0$ and using the fact that the irrelevant vertices at
$\Lambda_0$ are equal to zero, just as in part a) of the proof of
lemma 1, i) is seen to be true at the next order.

b) We now investigate the irrelevant vertices for $\Lambda \in
[\Lambda_m,\Lambda_M]$. If we integrate the left
hand side of \edivi\ from $\Lambda$ up to $\Lambda_M$, where $\Lambda_m
\leq \Lambda\leq \Lambda_M$, we obtain the equality
\eqn\edix{ \partial^j_p
V^{r,s,t}_{2m,2n}(\Lambda) -  \partial_p^j
V^{r,s,t}_{2m,2n}(\Lambda_M)=\quad
\int^{\Lambda_M}_{\Lambda} d \Lambda' {\partial \over \partial
\Lambda'} \biggl(  \partial^j_p
V^{r,s,t}_{2m,2n}(\Lambda')\biggr),}
which is easily turned into the inequality
\eqnn\edx
$$\Vert \partial^j_p
 V^{r,s,t}_{2m,2n}(\Lambda)\Vert_{\Lambda,\Lambda_M}
\leq \Vert  \partial^j_p V^{r,s,t}_{2m,2n}(\Lambda_M)
\Vert_{\Lambda_M,\Lambda_M} +
\int^{\Lambda_M}_{\Lambda} d \Lambda' \Biggl\Vert {\partial \over
\partial\Lambda'} \biggl(  \partial^j_p
V^{r,s,t}_{2m,2n}(\Lambda')\biggr)\Biggr\Vert_{\Lambda',\Lambda_M}.\edx$$
We first consider the vertices with $n>0$, in which case all vertices on the
right--hand side of \edivi\ also have $n>0$. Simply by evaluating i)
at $\Lambda=\Lambda_M$, we know that the first term on the right--hand
side of \edx\ is $\leq \Lambda_M^{4-2m-2n-j} P\log({\Lambda_M\over
\Lambda_m})$. However, we might
think that the bound on $V^{r,s,t}_{0,2}(\Lambda)$ and its first
momentum derivative would cause
the second term on the right of \edx\ to be inconsistent with
(ii). It actually causes no problem since, as we see from \edivi, it
always appears along with the factor
$K^{1/2}(M^2/\Lambda^2)$. $M \sim \Lambda_M$, so as far as the right
hand side of the flow
equation is concerned the bound on $V^{r,s,t}_{0,2}(\Lambda)$ is
\eqn\edci{\Vert V^{r,s,t}_{0,2}(\Lambda) \Vert_{\Lambda_M}
\leq \Lambda_M^2 \hbox{$K^{1/2}({\Lambda_M^2 \over  \Lambda^2})$}
P\log\biggl( {\Lambda_M
\over \Lambda_m}\biggr) \leq \Lambda^2 P \log\biggl( {\Lambda_M
\over \Lambda_m}\biggr),}
since $xK^{1/2}(x) \leq c$. A similar result is obviously true for the
first momentum derivative of $V^{r,s,t}_{0,2}(\Lambda)$. We can
therefore say that
\eqn\edl{\Biggl\Vert {\partial \over
\partial\Lambda} \biggl(  \partial^j_p
V^{r,s,t}_{2m,2n}(\Lambda)\biggr)\Biggr\Vert_{\Lambda,\Lambda_M} \leq
\Lambda^{3-2m-2n-j}P\log\biggl({\Lambda_M\over \Lambda_m}\biggr),}
for $n>0$. Thus, the term on the far right in \edx\ is
perfectly consistent with ii), and \edx\ coupled with the induction
hypothesis and the bounds
already obtained on the vertices at $\Lambda_M$ immediately lead to
the verification of ii) for all the irrelevant vertices with $n>0$.

We can now prove ii) for the irrelevant vertices with $n=0$. The
only obstacle to
proving this bound using \edx\ comes from
the term on the right--hand side of \edivi\ containing $\Vert
V^{r,s,t}_{2m,2}(\Lambda)\Vert_{\Lambda,\Lambda_M}$. From above we
already know that when $\Lambda$
is between $\Lambda_m$ and $\Lambda_M$, $\Vert
V^{r,s,t}_{2m,2}(\Lambda)\Vert_{\Lambda,\Lambda_M} \leq
\Lambda^{2-2m-j}P\log(\Lambda_M/\Lambda_m)$ for all vertices which could
appear on the right--hand side of \edivi. Thus,
within this range of
$\Lambda$, the problematic term is bounded by
$\Lambda^{3-2m-j}K^{1/2}({\Lambda_M^2 \over\Lambda^2})
P\log ({\Lambda_M \over \Lambda_m})$.
To show that this term in the flow equation is consistent with the induction
argument needed to prove ii), we may use the inequality
\eqn\edli{ \hbox{$K^{1/2}({\Lambda_M\over\Lambda})$}
P\Biggl[\log\biggl({\Lambda_M\over \Lambda}\biggr)+\log
\biggl({\Lambda\over \Lambda_m}\biggr)\Biggr] \leq P\log
\biggl({\Lambda \over \Lambda_m}\biggl),}
for all $\Lambda\in [\Lambda_m,\Lambda_M]$. By substituting this bound
into the right hand side of \edivi\ we obtain
\eqn\edlii{\Biggl\Vert {\partial \over
\partial\Lambda} \biggl( \partial^j_p
V^{r,s,t}_{2m,0}(\Lambda)\biggr)\Biggr\Vert_{\Lambda,\Lambda_M} \leq
\Lambda^{3-2m-j}\Biggl(P\log\biggl({\Lambda\over \Lambda_m}\biggr) +
{\Lambda \over \Lambda_0} P\log \biggl({\Lambda_0
\over\Lambda_m}\biggr)\Biggr),}
for all $\Lambda \in [\Lambda_m,\Lambda_M]$. The induction argument now goes
through as normal and the irrelevant vertices with $\phi$ legs only
satisfy ii).

c) The proofs of iii) for the irrelevant vertices follow much the
same lines as those of ii). In both cases we integrate the
left--hand side of \edivi\ from $\Lambda$ up to $\Lambda_m$. The
boundary conditions obtained from ii), evaluated at $\Lambda
= \Lambda_m$, are consistent with iii). We
therefore only have to be concerned with the right--hand side of
\edivi. We first consider $n>0$. We may substitute
$K^{1/2}({M^2\over \Lambda^2})\Lambda^{-3-j_1}$ in
the last term on the right--hand side by $\Lambda_m^{-3-j_1}$ (we
could obviously do better than this if we wanted, but this result is
sufficient to prove the lemma), or for the two point vertex
for the heavy field, replace
$K^{1/2}(M^2/\Lambda^2)\Lambda^{-3-j_1}\Lambda_M^2$ by
$\Lambda^{1-j_1}$. Once we have done this we may write
\eqn\edliii{\Biggl\Vert {\partial \over
\partial\Lambda} \biggl(  \partial^j_p
V^{r,s,t}_{2m,2n}(\Lambda)\biggr)\Biggr\Vert_{\Lambda_m,\Lambda_M} \leq
\Lambda_m^{3-2m-2n-j}P\log\biggl({\Lambda_M\over \Lambda_m}\biggr),}
for $n>0$ and all $\Lambda \in [0,\Lambda_m]$. With this result the
proof of iii) immediately goes through. For $n=0$, it is again only
the term containing
$\Vert V^{r,s,t}_{2m,2}(\Lambda)\Vert_{\Lambda_m,\Lambda_M}$ that
causes a potential obstruction. We easily see that
$K^{1/2}(\Lambda_M^2/\Lambda^2) P\log({\Lambda_M\over \Lambda_m}) \leq
c$ for $\Lambda \leq \Lambda_m$, and so
\eqn\edliv{\Biggl\Vert {\partial \over
\partial\Lambda} \biggl(  \partial^j_p
V^{r,s,t}_{2m,0}(\Lambda)\biggr)\Biggr\Vert_{\Lambda_m,\Lambda_M} \leq
\Lambda_m^{3-2m-j},}
for all $\Lambda \in [0,\Lambda_m]$. The proof of iii) for $n=0$ can
now also go through without obstruction.

\medskip

d) It is now possible to show that the lemma is also true for the
relevant vertices. We start with the four--point vertex with  $\Phi$
legs, and prove ii) and iii). In order to do this we use the
renormalization conditions \eIxv\ at
$\Lambda = 0$ to provide bounds on the relevant coupling constant,
i.e the vertex for the momenta $\tilde P_i$ at which the renormalization
conditions are set, at $\Lambda_M$. We therefore use \edivi, and the
bounds already obtained for the vertices at lower order in the
expansion parameters or
at equal order in the expansion parameters, but with greater numbers
of legs, to verify \edl\ and \edlii, and write
\eqn\edlv{\Biggl\vert {\partial \over \partial \Lambda}
V^{r,s,t}_{0,4}(\tilde P_i;\Lambda) \Biggr\vert \leq
\Lambda^{-1}P\log\biggl({\Lambda_M\over \Lambda_m}\biggr),}
for $\Lambda\in[\Lambda_m,\Lambda_R]$, and the same if we replace
$\Lambda^{-1}$ by $\Lambda_m^{-1}$ for $\Lambda\in[0,\Lambda_m]$. But,
\eqn\edlvi{\vert V^{r,s,t}_4(\tilde P_i;\Lambda_R) \vert
\leq \vert V^r_{0,4}(\tilde P_i;0) \vert
+ \int^{\Lambda_R}_{0} d
\Lambda' \Biggl\vert {\partial \over \partial
\Lambda'}V^{r,s,t}_{0,4}(\tilde P_i;\Lambda')\Biggr\vert.}
So, using the renormalization condition on $V^{r,s,t}_{0,4}(\tilde P_i;0)$, and
splitting the integral into two, one from $\Lambda=0$ to $\Lambda_m$
and the other from $\Lambda_m$ to $\Lambda_M$, we find that
\eqn\edlvii{\vert V^{r,s,t}_{0,4}(\tilde P_i;\Lambda_M)\vert
\leq P\log\biggl({\Lambda_R\over \Lambda_m}\biggr),}
and we obtain a bound on the vertex
defined at $\Lambda_M$ for the particular
momenta at which the renormalization condition is set. Using Taylor's
formula at $\Lambda =\Lambda_M$ we can verify
ii) for $\Lambda =\Lambda_M$, and obtain a boundary condition on
the vertex at $\Lambda_M$; $\Vert
V^{r,s,t}_{0,4}(\Lambda_M)\Vert_{\Lambda_M}\leq P\log({\Lambda_M\over
\Lambda_m})$. Using this boundary condition it is then straightforward to
verify ii) and iii) for $V^{r,s,t}_{0,4}(\Lambda)$ in the
same way as these bounds were verified for the irrelevant vertices.
Thus, ii) and iii) are verified for the four--point vertex with
$\Phi$ legs only.

Exactly the same argument works for $V^{r,s,t}_{2,2}(\Lambda)$ and for
$\partial^2_pV^{r,s,t}_{0,2}(\Lambda)$. We may then verify iv) for
$\Vert \partial_pV^{r,s,t}_{0,2}(\Lambda)\Vert_{\Lambda_M}$ by using the bounds
on
$\Vert \partial^2_pV^{r,s,t}_{0,2}(\Lambda)\Vert_{\Lambda_M}$ and Taylor's
formula about zero
momentum. It is then straightforward to verify iv) for
$\Vert V^{r,s,t}_{0,2}(0)\Vert_{\Lambda_M}$ by using the boundary condition on
the vertex at
$\Lambda_0$ and Taylor's formula, obtaining $\Vert
V^{r,s,t}_{0,2}(0)\Vert_{\Lambda_M}
\leq \Lambda_M^2 P\log({\Lambda_M\over \Lambda_m})$. Using this
boundary condition on the vertex we may verify iv) for all $\Lambda
\in [0,\Lambda_m]$ by integrating the left--hand side of \edivi\ from
$0$ to $\Lambda$ and using \edlii. Also, using the boundary condition from
evaluating (vi) at $\Lambda_m$, we then verify iv) for all $\Lambda
\in [\Lambda_m, \Lambda_M]$ by integrating \edivi\ from $\Lambda_m$
up to $\Lambda$ and using \edl. In this way we may verify ii), iii) and
iv) for all the relevant vertices with $n>0$.

We can then prove iii)
for the relevant vertices involving only $\phi$
fields. The method for doing this is essentially the same as in
\REFT. Once we use the bounds obtained from iii) in order to bound the term
involving $\Vert V^{r,s,t}_{2m,2}(\Lambda)\Vert_{\Lambda,\Lambda_M}$, the
right-hand side of \edivi\ is consistent with iii), i.e we obtain the
result in \edliv. Thus, it simply remains to verify i) for
the relevant vertices, and ii) with $n=0$.

e) In order to prove ii) for the relevant vertices with $n=0$ we
use the same method as used in part d) of lemma 1, i.e.
we integrate the left--hand side of \edivi\ defined at the momenta at which the
renormalization conditions were set from $\Lambda\in
[\Lambda_m,\Lambda_M]$ down to $\Lambda_m$, and use the derived
boundary condition on the coupling constant at $\Lambda_m$ (obtained
from iii) evaluated at $\Lambda_m$ for the particular momenta) to
obtain a bound on the coupling constant defined at $\Lambda$. For
example, we find that $\vert V^{r,s,t}_{4,0}(P_i;\Lambda)\vert \leq P
\log ({\Lambda \over \Lambda_m}) + {\Lambda \over
\Lambda_0}P\log({\Lambda_m\over \Lambda_0})$. We may then verify ii) by
using these results along with Taylor's formula. We simply have to
remember to work downwards in number of legs, and then in number of
derivatives.

f) Finally, we may verify i) for all the relevant vertices in the same
manner. We simply integrate the $\Lambda$--derivative of the coupling
constants from $\Lambda$ down to $\Lambda_M$, use the derived
boundary conditions on these couplings at $\Lambda_M$, along with the
bound on the $\Lambda$--derivative, to provide a bound on the coupling
constant defined at $\Lambda$, and use Taylor's formula to derive the
bound on the norm of the vertex. Again, we must do this first for the
four-point vertices, and then for the two-point vertices in decreasing
number of derivatives (though for given number of legs it does not
matter whether we deal with vertices with $\phi$ or $\Phi$ legs
first). Once this is complete we have verified i)--iv) for all
vertices.

\medskip

We have thus demonstrated
that lemma 8 is true for all $m$ and $n$ at orders $r$, $s$
and $t$ in the expansion coefficients. At next order in any of
these coefficients the lemma is true at large enough $m$ and $n$,
and again the proof by induction goes through and the lemma is true at
this next order. But the lemma is trivially satisfied at zeroth order
in all the expansion coefficients, and thus, by induction, lemma 8 is true for
all $m$, $n$, $r$, $s$ and $t$. \blackbox

\medskip

We have therefore been able to obtain a bound on all the vertices for
all $\Lambda$ up to, and including, $\Lambda_0$. In particular,
iii) and iv) evaluated at $\Lambda=0$ tell us that for $n > 0$, for all $m$,
$j$, except for $m=0$, $n=1$, and $j=0,1$,
\eqn\eliix{ \Vert \partial^j_p \tilde
G^{c}_{2m,2n}(\Lambda_0,\lambda_i)\Vert_{\Lambda_m,\Lambda_M} \leq
\Lambda_m^{4-2m-2n-j} P\log \biggl({\Lambda_M \over \Lambda_m}\biggr);}
for $n=0$, for all $m$, $j$,
\eqn\edlix{ \Vert \partial^j_p \tilde
G^{c}_{2m,0}(\Lambda_0,\lambda_i)\Vert_{\Lambda_m} \leq
\Lambda_m^{4-2m-j};}
and for the special case $m=0$, $n=1$, $j=0,1$, the heavy two--point vertex
satisfies
\eqn\edlx{ \Vert \partial^j_p \tilde G^{c}_{0,2}(\Lambda_0,\lambda_i)
\Vert
_{\Lambda_M}\leq \Lambda_M^{2-j} P \log\biggl( {\Lambda_M
\over \Lambda_m}\biggr).}
These bounds show that the amputated connected Green's functions
have a maximum
value which is independent of $\Lambda_0$, and thus that they remain
finite in the limit $\Lambda_0\to\infty$.

\subsec{Convergence and Universality.}

By combining the methods used to prove the convergence and
universality of Green's functions in the infinite cut--off limit
in \REFT\ (lemma 2 and lemma 3 respectively)  with those
described above to prove lemma 8, it is now relatively straightforward to
prove convergence and universality lemmas for the theory
under consideration here. The proofs of these lemmas will thus be
omitted.

The convergence lemma takes the form

\medskip
\noindent {\it Lemma 9:}\nobreak

\noindent For $n>0$,, for all $m, j,$ except for $m =0, n=1$, and $j=0,1$,
\eqn\edlxi{\Biggl\Vert \biggl(\Lambda_0{\partial \over \partial \Lambda_0}
\partial^j_p
\tilde G^{c}_{2m,2n}(\Lambda_0,\lambda_i) \biggr)_{{\lambda_i}}
\Biggr\Vert_{\Lambda_m,\Lambda_M}
\leq\biggl( {\Lambda_M\over\Lambda_0}\biggr)^2 \Lambda_m^{4-2m-2n-j}
 P\log \biggl({\Lambda_M \over\Lambda_m}\biggr)
P\log \biggl( {\Lambda_0 \over\Lambda_m}\biggr),}
while for $n=0$, for all $m,j$,
\eqn\edlxiii{\Biggl\Vert \biggl(\Lambda_0{\partial\over\partial\Lambda_0}
\partial^j_p
\tilde G^{c}_{2m,0}(\Lambda_0,\lambda_i) \biggr)_{{\lambda_i}}
\Biggr\Vert_{\Lambda_m} \leq
\biggl({\Lambda_m\over\Lambda_0}\biggr)^2 \Lambda_m^{4-2m-j}
P\log\biggl({\Lambda_0\over\Lambda_m}\biggr),}
and for the special case $m=0$, $n=1$, $j=0,1$,
\eqn\edlxc{ \Biggl\Vert \biggl(\Lambda_0{\partial \over \partial\Lambda_0}
\partial^j_p \tilde G^{c}_{0,2}(\Lambda_0,\lambda_i)\biggr)_{\lambda_i}
\Biggr\Vert_{\Lambda_M}\leq \biggl({\Lambda_M \over \Lambda_0}\biggr)^2
\Lambda_M^{2-j} P \log\biggl( {\Lambda_M
\over \Lambda_m}\biggr)P\log\biggl({\Lambda_0\over \Lambda_m}\biggr).}
Thus the connected amputated Green's
functions, and hence the S-matrix elements, calculated using the effective
Lagrangian at any scale between $\Lambda_m$ and $\Lambda_M$ have a
well defined limit as $\Lambda_0\rightarrow\infty$. Therefore
our theory containing
particles of masses $m$ and $M$ is
perturbatively renormalizable in the conventional sense.

To prove universality we consider, just as in lemma 3,
a second theory with more general
irrelevant boundary conditions at $\Lambda_0$ than those of the
first, \exxxv, namely
\eqn\elxxvi{\eqalign{\bar V^{r,s,t}_{2m,2n}(p,-p;\Lambda_0) =
\bar\lambda^{r,s,t}_{2m,2n}(\Lambda_0) \Lambda_0^2 +
p^2 \bar\lambda^{r,s,t}_{2m,2n}(\Lambda_0)
+ \Lambda_0^2 \eta^{r,s,t}_{2m,2n}(p,-p;\Lambda_0),\qquad
& \hbox{$m+n=1$},\cr
\bar V^{r,s,t}_{2m,2n}(p_1,p_2,p_3,p_4;\Lambda_0) =
\bar\lambda^{r,s,t}_{2m,2n}(\Lambda_0)
+ \eta^{r,s,t}_{2m,2n}(p_1,p_2,p_3,p_4;\Lambda_0),\qquad
&\hbox{$m+n=2$},\cr
\bar V^{r,s,t}_{2m,2n}(p_1,\cdots,p_{2(m+n)};\Lambda_0) =
\Lambda_0^{4-2(m+n)}
\eta^{r,s,t}_{2m,2n}(p_1,\cdots,p_{2(m+n)};\Lambda_0),\qquad
&\hbox{$m+n>2$}.\cr}}
The functions $\eta^{r,s,t}_{2m,2n}(p_i;\Lambda_0)$ satisfy the same
general conditions as those in \S 3.1 of \REFT, namely they are real functions
of the momenta $p_i$, regular when continued into the
complex plane, are natural in the sense that
\eqn\elxxix{ \Vert \partial^j_p \eta^{r,s,t}_{2m,2n}(\Lambda_0)
\Vert_{\Lambda_0}\leq
\Lambda_0^{-j} P \log \biggl( {\Lambda_0 \over \Lambda_M} \biggr)}
for $2(m+n)+j>4$, and vanish for $m+n>r+s+t+1$.
By convention we also take $\eta^{r,s,t}_{2m,2n}(\Lambda_0)=0$ if
$m=0$ and $r>0$, or if $n=0$ and $s>0$, or if $mn=0$ and $rs>0$.

We can then show that
if we write $\Delta\tilde G^{c}_{2m,2n}$ for the difference between
the amputated connected Green's functions of the two theories,

\medskip
\noindent {\it Lemma 10:}\nobreak

\noindent For $n>0$, for all $m, j$, except for $m=0, n=1$, and $j=0,1$
\eqn\edlxie{\bigl\Vert
\partial^j_p
\Delta\tilde G^{c}_{2m,2n}
\bigr\Vert_{\Lambda_m,\Lambda_M}
\leq\biggl( {\Lambda_M\over\Lambda_0}\biggr)^2 \Lambda_m^{4-2m-2n-j}
 P\log \biggl({\Lambda_M \over\Lambda_m}\biggr)
P\log \biggl( {\Lambda_0 \over\Lambda_m}\biggr),}
while for $n=0$, for all $m, j$,
\eqn\edlxiiie{\bigl\Vert
\partial^j_p\Delta\tilde G^{c}_{2m,0}
\bigr\Vert_{\Lambda_m} \leq
\biggl({\Lambda_m\over\Lambda_0}\biggr)^2
\Lambda_m^{4-2m-j}
P\log\biggl({\Lambda_0\over\Lambda_m}\biggr),}
and for the special case $m=0$, $n=1$, $j=0,1$,
\eqn\edlxu{ \Vert \partial^j_p \Delta\tilde G^{c}_{0,2}
\Vert
_{\Lambda_M}\leq \biggl({\Lambda_M\over \Lambda_0}\biggr)^2\Lambda_M^{2-j} P
\log\biggl( {\Lambda_M
\over \Lambda_m}\biggr)P\log\biggl({\Lambda_0 \over \Lambda_m}\biggr).}
Thus all S--matrix elements are universal, in the sense that they are
independent, up to power suppressed terms, of all the irrelevant
parameters $\eta^{r,s,t}_{2m,2n}(p_i;\Lambda_0)$, provided that these
are natural at $\Lambda_0$.
The effective theory of the two particles is thus renormalizable
in the precise sense
explained in \REFT. As expected from power counting, corrections to
light particle processes are suppressed $\Lambda_m^2/\Lambda_0^2$
(up to logarithms), while those involving heavy particles are
only suppressed by powers of $\Lambda_M^2/\Lambda_0^2$.

Both lemma 9 and lemma 10 may be systematically improved, after the
fashion of \S 3.2 of \REFT. We here defer discussion of systematic
improvement until \S 4.

\subsec{Weinberg Bounds.}

The bound \eliix\ derived from lemma 8 is actually rather weak; by
considering in more detail the number of heavy particle
legs on a given vertex, rather than just splitting vertices into those
with either some heavy particle legs or no heavy particle legs, it is possible
to obtain much more stringent bounds on the Green's functions with
heavy particle legs (we
already have the best possible bound, namely \edlix, for Green's
functions with light particle legs only). To do this we need to also
incorporate the notation and arguments in ref.\Infrared\ in
order to consider the number $\tilde e$ of heavy particle exceptional
momenta for a particular vertex (i.e. the total number of momenta within
all irreducible sets (except for the largest) which have magnitudes
between some minimum value $E$ and $\Lambda_M$). The resulting
bounds, analogous to lemma 6, are summarized in the following lemma:

\medskip
\noindent {\it Lemma 11:}\nobreak

\noindent i) For all $m>0$, $n>1$, and $0\leq \tilde{e}\leq 2n-3$
\eqn\eDd{\bigl\Vert \partial^j_p\tilde G^{c}_{2m,2n}(\Lambda_0,\lambda_i)
\bigr\Vert_{\Lambda_m,\Lambda_M}^{E,\tilde{e}} \cases
{\leq\Lambda_M^{3+\tilde{e}-2n}\Lambda_m^{2-2m-j}\bar\Lambda^{-1-\tilde{e}}
P\log \Bigl({\Lambda_M \over \Lambda_m}\Bigr) & $\tilde{e}$ odd,\cr
\leq\Lambda_M^{2+\tilde{e}-2n}\Lambda_m^{2-2m-j}\bar\Lambda^{-\tilde{e}}
P\log \Bigl({\Lambda_M\over\Lambda_m}\Bigr) & $\tilde{e}$ even,\cr}}
where $\bar\Lambda={\rm max}(E,\Lambda_m)$;

\noindent ii) for  $m=0$, and $n>0$, except for $n=1, j=0,1$, then for
$\tilde{e}=0$
\eqn\eDe{ \bigl\Vert\partial^j_p\tilde G^{c}_{0,2n}(\Lambda_0,\lambda_i)
\bigr\Vert_{\Lambda_m,\Lambda_M}\leq\Lambda_M^{4-2n-j};}
\noindent while for $1\leq \tilde{e}\leq 2n-3$,
\eqn\eDf{\bigl\Vert \partial^j_p \tilde G^{c}_{0,2n}(\Lambda_0,\lambda_i)
\bigr\Vert_{\Lambda_m,\Lambda_M}^{E,\tilde{e}} \cases{
\leq\Lambda_M^{3+\tilde{e}-2n}\bar\Lambda^{1-\tilde{e}-j}
P\log \Bigl({\Lambda_M \over \bar \Lambda}\Bigr) & $\tilde{e}$ odd,\cr
\leq\Lambda_M^{2+\tilde{e}-2n}\bar\Lambda^{2-\tilde{e}-j}
P\log \Bigl({\Lambda_M\over \bar \Lambda}\Bigr) &$\tilde{e}$ even;\cr}}
\noindent iii) for $m=0$, $n=1$, $j=0,1$,
\eqn\edlxw{ \Vert \partial^j_p \tilde G^{c}_{0,2}(\Lambda_0,\lambda_i)
\Vert_{\Lambda_M}\leq \Lambda_M^{2-j};}
\noindent iv) for all $m>0$, $n=1$,
\eqn\eDb{\bigl\Vert\partial^j_p\tilde G^{c}_{2m,2}(\Lambda_0,\lambda_i)
\bigr\Vert_{\Lambda_m,\Lambda_M}
\leq\Lambda_m^{4-2m-j}P\log\biggl({\Lambda_M\over\Lambda_m}\biggr).}
Here $\tilde{e}$ denotes the maximum number of exceptional momenta
for the external
legs corresponding to heavy particles; all the external momenta for
light particle legs can be individually exceptional as low
as zero. All derivatives are assumed to act on light particle legs, or
else heavy particle legs carrying exceptional momenta (if this is not
the case they would produce inverse powers of $\Lambda_M$ rather than
$\bar \Lambda$). The vertices with $m>0$ and $n=1$ are a special case
because the large number of light particle legs mean that the total
number of possible exceptional momenta already exceeds $2m +2n -3$.
It is convenient to have these best possible bounds, since then the bounds
are comparable to the renormalization conditions on the
vertices involving heavy particles.\foot{Though
in practice the on--shell renormalization conditions for
vertices with heavy particle legs may be larger than the
corresponding Euclidean
space bounds because of the proximity of the poles in the Green's
functions to branch points due to light particle bremmstrahlung.}
However as lemma 8 is sufficient
to prove the decoupling theorem,
we leave the proof of lemma 11 (a straightforward, but rather lengthy,
combination of the methods of this paper and \Infrared) to the more
enthusiastic reader.

\newsec{The Decoupling Theorem.}

In the context of the effective scalar field theories discussed so
far, the decoupling theorem consists of the following statement:
if we calculate connected amputated Green's functions involving
only light particles at scales $\sim \Lambda_m$, using an
effective theory containing only $\phi$ fields (as defined in \REFT),
with naturalness scale $\Lambda_M$, then these Green's functions
will be the same as those calculated using the full theory (as defined
in \S 1 above), up to corrections suppressed by powers of
$(\Lambda_m/\Lambda_M)$.

When viewed in this way, decoupling is nothing but a generalization of
universality: the light particle theory with regular boundary
conditions on its irrelevant vertices is approximately equivalent at
low energies to one in which these vertices have heavy particle poles
and cuts. Indeed the method we will use to prove the decoupling
theorem is very similar to
that used to prove universality in \S 3.1 of \REFT.

In this section we state and prove a lemma which contains the
decoupling theorem in its most simple form. Then in the next section we
will generalize this result. We thus introduce a theory
containing only $\phi$ fields, and in fact defined precisely as in \S
2.1 of \REFT, with zero boundary conditions on the irrelevant couplings
at $\Lambda_M$, and boundary conditions on the relevant couplings
which are the same as \eIxv\ for the vertices containing only
$\phi$ fields. We denote the vertices of this new theory by
$\hat V^{r,s,t}_{2m}(p_1.....p_{2m};\Lambda)$, and consider the difference
between this theory and the one containing both $\phi$ and $\Phi$
fields by introducing the quantity
\eqn\edxxiv{D^{r,s,t}_{2m}(\Lambda) \equiv
 V^{r,s,t}_{2m,0}(\Lambda) - \hat V^{r,s,t}_{2m}(\Lambda),}
which gives us a measure of the difference between the two theories.

Subtracting the flow equation for the $\hat V^{r,s,t}_{2m}(\Lambda)$ away
from that for the $V^{r,s,t}_{2m,0}(\Lambda)$ and taking norms, we
easily obtain, for $\Lambda \in [0,\Lambda_M]$,
\eqn\edxxv{\eqalign{\Biggl\Vert &{\partial \over \partial \Lambda}\Bigl(
\partial^j_p D^{r,s,t}_{2m}(\Lambda)\Bigr) \Biggr\Vert_{\Lambda'}
\leq \Lambda\Vert \partial^j_p D^{r,s,t}_{2m+2}(\Lambda)\Vert_{\Lambda'}
+ \hbox{$K^{1/2}({\Lambda_M^2\over \Lambda^2})$}
   \Lambda\Vert\partial^j_p V^{r,s,t}_{2m,2}(\Lambda)
               \Vert_{\Lambda',\Lambda_M}\cr
&+ \sum^{m}_{l=1}\sum^{r-1}_{r'=1}\sum_{s'=1}^{s-1}\sum_{t'=1}^{t-1}
   \sum_{j_i;j_1 +j_2 + j_3 = j} \Lambda^{-3-j_1}
\big[\Vert\partial^{j_2}_{p}D^{r',s',t'}_{2l}(\Lambda)\Vert_{\Lambda'}\cr
& \hskip 2.5in +\Vert\partial^{j_2}_{p}\hat
V^{r',s',t'}_{2l}(\Lambda)\Vert_{\Lambda'}\big]
\cdot\Vert\partial^{j_3}_p
D^{r-r',s-s',t-t'}_{2m+2-2l}(\Lambda)\Vert_{\Lambda'} \cr}}
where again $\Lambda'={\rm max}(\Lambda,\Lambda_m)$ just as in \edivi.

If we consider $\Lambda\in[\Lambda_m,\Lambda_M]$, we have the equality
\eqn\edxxvi{   \partial^j_p
D^{r,s,t}_{2m}(\Lambda) -  \partial_p^j
D^{r,s,t}_{2m}(\Lambda_M) =
\int^{\Lambda_M}_{\Lambda} d \Lambda' {\partial \over \partial
\Lambda'} \biggl(  \partial^j_p
D^{r,s,t}_{2m}(\Lambda')\biggr),}
which in the same way as previous equalities is easily turned into
the inequality
\eqn\edxxvii{\Vert  \partial^j_p D^{r,s,t}_{2m}(\Lambda)\Vert_{\Lambda}
\leq \Vert  \partial^j_p D^{r,s,t}_{2m}(\Lambda_M)\Vert_{\Lambda_M} +
\int^{\Lambda_M}_{\Lambda} d \Lambda' \Biggl\Vert {\partial \over
\partial\Lambda'} \biggl( \partial^j_p
D^{r,s,t}_{2m}(\Lambda')\biggr)\Biggr\Vert_{\Lambda'}.}
This equation is useful for bounding the difference between the
irrelevant vertices for our two theories for $\Lambda \in
[\Lambda_m,\Lambda_M]$. For $\Lambda \in [0,\Lambda_m]$ we can derive
similarly
\eqn\edxxviim{\Vert  \partial^j_p D^{r,s,t}_{2m}(\Lambda)\Vert_{\Lambda_m}
\leq \Vert  \partial^j_p D^{r,s,t}_{2m}(\Lambda_m)\Vert_{\Lambda_m} +
\int^{\Lambda_m}_{\Lambda} d \Lambda' \Biggl\Vert {\partial \over
\partial\Lambda'} \biggl( \partial^j_p
D^{r,s,t}_{2m}(\Lambda')\biggr)\Biggr\Vert_{\Lambda_m}.}

To obtain equations which are useful for finding bounds
on the difference between
the relevant coupling constants in the range $\Lambda \in
[\Lambda_m,\Lambda_M]$ it is necessary to integrate with
respect to $\Lambda$ down to $\Lambda = \Lambda_m$, put the momenta
equal to those at which the renormalization conditions on the relevant coupling
constants
are set, and take bounds to
obtain
\eqn\edxxviii{ \Bigl\vert \partial^j_p
D^{r,s,t}_{2m}(\Lambda)\vert_{p_i=P_i}\Bigr\vert \leq \Bigl\vert  \partial^j_p
D^{r,s,t}_{2m}(\Lambda_m)\vert_{p_i=P_i}\Bigr\vert +
\int^{\Lambda}_{\Lambda_m} d\Lambda'
\Biggl\Vert {\partial \over \partial \Lambda'} \partial^j_p
D^{r,s,t}_{2m}(\Lambda')\Biggr\Vert_{\Lambda'}.}
For an equation useful for finding bounds on the difference between
the relevant vertices for our two theories for $\Lambda \in
[0,\Lambda_m]$, we must integrate from $\Lambda$ down to $0$ and take
bounds with respect to $\Lambda_m$ to obtain
\eqn\edxxviiim{ \Vert \partial^j_p
D^{r,s,t}_{2m}(\Lambda)\Vert_{\Lambda_m} \leq \Vert  \partial^j_p
D^{r,s,t}_{2m}(\Lambda_m)\Vert_{\Lambda_m} +
\int^{\Lambda}_{0} d\Lambda'
\Biggl\Vert {\partial \over \partial \Lambda'} \partial^j_p
D^{r,s,t}_{2m}(\Lambda') \Biggr\Vert_{\Lambda_m}.}

Equations \edxxv\ --\edxxviiim, together with the values of the
vertices at $\Lambda_M$ and the bounds on the vertices in both theories, which
are already known from lemmas 1 and 5
(since the theory containing just the $\phi$ fields is the same as
that in \REFT\ with $\Lambda_m =\Lambda_R$, $\Lambda_M$ instead of
$\Lambda_0$), will now be shown to lead to a bound on the
difference between the vertices, and thus the Green's functions, in
the two theories. These bounds are summarized in
\medskip
\noindent {\it Lemma 12:}\nobreak

\noindent i) For all $\Lambda \in [\Lambda_m, \Lambda_M]$,
\eqn\edxxix{\Vert \partial^j_p D^{r,s,t}_{2m}(\Lambda)\Vert \leq {\Lambda
\over \Lambda_M} \Lambda^{4-2m-j} P \log \biggl( {\Lambda_M \over
\Lambda_m} \biggr).}
\noindent ii) For all $\Lambda \in [0, \Lambda_m]$,
\eqn\edxxixa{\Vert \partial^j_p D^{r,s,t}_{2m}(\Lambda)\Vert_{\Lambda_m} \leq
{\Lambda_m
\over \Lambda_M} \Lambda_m^{4-2m-j} P \log \biggl( {\Lambda_M \over
\Lambda_m} \biggr).}

\medskip

Once again, the proof follows the same induction scheme as that
used to prove lemma 1; we assume that the lemma is true at order $r-1,s-1,t-1$
in $g_m$, $g_M$ and $g_{mM}$, and vertices with more than $2m+2$ legs at next
order in the coupling constants, and then show that it remains
true for vertices with $2m$ legs at order $r,s,t$. Each induction step
then follows the same four steps a)--d) as lemma 1.

a) For the irrelevant vertices with $\Lambda \in
[\Lambda_m,\Lambda_M]$ we simply have to
use \edxxvii\ and the fact that the values of the vertices in each theory at
$\Lambda_M$ are consistent with lemma 12, and therefore so is their
difference, as we see by substituting $\Lambda =\Lambda_M$ into lemma
8ii), and i) is immediately seen to be true at this order in the
expansion parameters for vertices
with $2m$ legs. The second term in \edxxv\ causes no problem since,
using \edvi\ we see that it is bounded by
\hbox{$K^{1/2}({M^2 \over \Lambda^2})\Lambda^{3-2m-j}
P\log ({\Lambda_M \over \Lambda_m})$},
which is less than or equal to the bound
\hbox{$\Lambda^{3-2m-j}({\Lambda\over\Lambda_M})
P\log ({\Lambda_M \over \Lambda_m})$}
required of all terms on the right-hand side of \edxxv\ in
order for the induction argument used to prove i) to go through.

b) The proof of ii) for the irrelevant vertices is much the same. We
use the derived boundary conditions obtained by evaluating i) at
$\Lambda_m$ along with \edxxviim, and ii) is clearly verified.
As in a) the second term in \edxxv\ is no obstruction since,
using  \eddvi\ and the fact that the vertex has an associated factor
of $K^{1/2}({\Lambda_M^2\over\Lambda^2})$, we see that it is bounded by
$\Lambda_m^{3-2m-j}({\Lambda_m\over \Lambda_M})P\log({\Lambda_M\over
\Lambda_m})$.

c) The difference between the relevant coupling constants in the two
theories is zero by definition. Using this result along with Taylor's
formula and the
bounds already obtained for the irrelevant vertices we can prove ii)
for each of the relevant vertices, by working down in number of legs
and number of derivatives. In order to prove ii) for the relevant
vertices for all $\Lambda \in [0,\Lambda_m]$ we use the derived
boundary conditions at $\Lambda=0$ along with \edxxviiim, again working
downwards in number of legs and number of derivatives. The potentially
troublesome second term in \edxxv\ is dealt with in the same way as in
b).

d) From ii) there are now good bounds on the difference between the relevant
coupling constants at $\Lambda_m$. Feeding these into \edxxviii\ it is
simply a repetition of previous exercises to bound the relevant
couplings and to prove
the lemma for the relevant vertices, using Taylor's formula and the
same methods used for the relevant vertices in \S 's 3.1 and 3.2.
Once this is done, then as explained in these previous sections, the
proof by induction is complete, and lemma 12 is true for all $r$, $s$,
$t$ and $m$. \blackbox

\medskip

In particular, setting $\Lambda = 0$, lemma 12 yields
\eqn\edxxxi{\Vert \partial^j_p (
\tilde G^{c}_{2m,0}(\Lambda_0,\lambda_i) -
{\hat G}^{c}_{2m}(\Lambda_M,\lambda_i)) \Vert_{\Lambda_m} \leq
{\Lambda_m \over \Lambda_M}
\Lambda_m^{4-2m-j} P \log \biggl({\Lambda_M \over \Lambda_m}\biggr),}
where the amputated connected Green's functions for the theory with
both light and heavy particles $\tilde G^{c}_{2m,0}$ depend on
$\lambda_i$, the seven coupling constants corresponding to the relevant
vertices with either (or both) light and heavy particle legs,
while the amputated
connected Green's functions for the light particle theory,
${\hat G}^{c}_{2m}$, depend only on the three coupling constants
corresponding to the relevant vertices with light particle legs only. Thus, we
see that if we simply delete the heavy particle fields $\Phi$ from the original
theory, while keeping the renormalization conditions on the relevant
couplings at $\Lambda = 0$ fixed, we change the
amputated connected Green's functions including only
$\phi$ fields at external momenta of order $\Lambda_m$ by
terms of order $\Lambda_m/\Lambda_M$. This is particularly transparent
if we adopt the convention described following \eIxv. In this
case the vertices in the theory containing only the light particle
have an expansion in $g_m$ only, and we see from \edxxixa\ that all
terms in the low--energy light particle vertices in the full theory at
greater than zeroth order in either $g_M$ or $g_{mM}$ are suppressed by
${\Lambda_m/\Lambda_M}$.

Thus, if we were to take $M/m\to\infty$ lemma 12 would therefore amount to
a proof of the conventional decoupling theorem\rdi. Combining
lemma 12 with the universality of both two theories (lemmas 3 and 10)
also gives us trivially
a proof of decoupling in effective theories. More subtly,
combining it with the proof of infrared finiteness of the light
particle theory in \Infrared\ (lemma 5) gives us a proof of the
decoupling theorem when the light particles are massless; this is very
difficult using more conventional techniques\ja.

\newsec{Systematic Improvement.}

Since in any realistic scenario the mass of the heavy particle is not
infinite but finite, it would be useful to improve our decoupling theorem in
much the same way that we improved the renormalizability of the
effective theory in \S 3.2 of \REFT. There we were able to show that
one can decrease the dependence of low energy Green's functions on the
irrelevant couplings at the naturalness scale $\Lambda_0$
by specifying more and more low energy renormalization conditions, and thus
determining the coupling constants corresponding to higher and higher
dimension operators. Here we might expect that it would be possible to
decrease the dependence of the Green's functions for the low mass
particle on the details of the theory of large mass particle $M$ by
specifying more and more renormalization conditions on the
theory containing only the light particle. It is easy to verify that this
is indeed the case.

First we consider the theory containing both particles. Consider
setting renormalization conditions on all irrelevant vertices up to a
given canonical dimension ${\tilde D} $. As with
\eIxv, these must be set with light particle legs having  momenta of
order $\Lambda_m$, and heavy particle legs with momenta of order
$\Lambda_M$. We can apply the same arguments as used in \REFT\ to
show that the bounds in lemma 8 are still satisfied, while those in
lemmas 9 and 10 are improved: \edlxi, \edlxc, \edlxie, and \edlxu\ all
acquire an
extra factor of $({\Lambda_M\over\Lambda_0})^{{\tilde D} -4}$ on the right hand
side,
while \edlxiii\ and \edlxiiie\ acquire an extra factor of
$({\Lambda_m\over\Lambda_0})^{{\tilde D} -4}$.

To improve the decoupling theorem, we repeat the argument in the
previous section, with the physically relevant vertices in the light
particle theory (all vertices with dimension not exceeding $D$,
where for the moment we let $D={\tilde D} $) having
identical low energy renormalization conditions to the vertices
involving light particles alone in the theory containing both
particle. For simplicity the remaining irrelevant vertices
(with dimension greater than $D$) are set to zero at $\Lambda_M$ in
the light particle theory, as are the undetermined irrelevant vertices
at $\Lambda_0$ in the full theory. Lemma 12 is then
superseded by the following improved bounds:

\medskip
\noindent {\it Lemma 13:}\nobreak

\noindent (i) For all $\Lambda \in [\Lambda_m,\Lambda_M]$,
\eqn\edxxxii{\Vert \partial^j_p D^{r,s,t}_{2m}(\Lambda) \Vert \leq
\biggl({\Lambda \over \Lambda_M}\biggr)^{D-2} \Lambda^{4-2m-j}P\log
\biggl( {\Lambda_M \over \Lambda_m}\biggr).}

\noindent (ii) For all $\Lambda \in [0,\Lambda_m]$,
\eqn\edxxxiim{\Vert \partial^j_p D^{r,s,t}_{2m}(\Lambda) \Vert_{\Lambda_m} \leq
\biggl({\Lambda_m \over \Lambda_M}\biggr)^{D-2} \Lambda_m^{4-2m-j}P\log
\biggl( {\Lambda_M \over \Lambda_m}\biggr).}

In order to prove this lemma we may use exactly the same induction
argument as that for lemma 12. Since both sets of boundary conditions on
$\partial^j_p D^{r,s,t}_{2m}(\Lambda)$ are consistent with the lemma,
and the flow equation \edxxv\ is also consistent with it (partly due to the
exponential in the second term on the right falling of more quickly than
any finite power), the proof goes through just as before, and we see
no reason to write it out in detail.
Just as for lemma 12 the assumptions on the irrelevant vertices may be
relaxed using the (improved) universality lemmas 4 and 10.

Setting $\Lambda= 0$ in \edxxxiim, we see immediately that \edxxxi\
now becomes
\eqn\edxxxie{\Vert \partial^j_p (
\tilde G^{c}_{2m,0}(\Lambda_0,\lambda_i) -
{\hat G}^{c}_{2m}(\Lambda_M,\lambda_i)) \Vert_{\Lambda_m} \leq
\biggl({\Lambda_m \over \Lambda_M}\biggr)^{D-2}
\Lambda_m^{4-2m-j} P \log \biggl({\Lambda_M \over \Lambda_m}\biggr);}
the amputated connected Green's functions for the two theories only
differ by terms of order $({\Lambda_m \over \Lambda_M})^{D-2}$ (up
to logarithms) when we consider external momenta with magnitudes of
order $\Lambda_m$.
So setting renormalization conditions on light particle couplings down
to dimension $4-D$ means
that we can calculate amputated connected Green's functions (and thus
S--matrix elements) for scatterings of light particles with energies
of order $\Lambda_m$ simply by using a
theory containing just the light particle field with precision of order
$({\Lambda_m\over\Lambda_M})^{D-2}$. Thus, as far as the bounding arguments
are concerned, we have shown that the naturalness scale
$\Lambda_0$ in \REFT\ could equally be the mass of another particle:
all the results proven in \S 2 and \S 3 of this paper sections being
also true in this case, as we might naively have expected.

However the physically relevant renormalization conditions for the
light particles are now dictated, up to small corrections, by
matching to the theory with the heavy particle: more precisely in
the phenomenologically `neutral' effective theory discussed in \REFT,
if all renormalization conditions corresponding to vertices of
dimension $D$ have been fixed, those of dimension $D+2$ may be
specified within a freedom of order $({\Lambda_m\over\Lambda_M})^{D-2}$
(as explained in \S 3.3 of \REFT); matching to the massive theory
means that this freedom is only of order
$({\Lambda_m\over\Lambda_0})^{D-2}$. This means, in fact, that we
have the freedom to choose $D$ to be greater than ${\tilde D} $, as long as
it satisfies the requirement that $({\Lambda_m\over\Lambda_0})^{{\tilde D} -2}
\leq ({\Lambda_m\over\Lambda_M})^{D-2}$. In this case we set the
renormalization conditions for the light particle theory to be
identical to those for the light particle vertices in the full theory
for dimension up to ${\tilde D} $ and equal to the values obtained by
calculating using the full theory for those vertices of higher
dimension. (If the renormalization conditions are set by matching to
experiment this will be true automatically, up to corrections of order
$({\Lambda_m\over \Lambda_0})^{{\tilde D} -2}$.) The proof of lemma 13 then
goes through with no obstruction and the conclusions expressed in
\edxxxie\ and the following paragraph still hold.

In particular, in the local limit ${\Lambda_m\over\Lambda_0}\to 0$, with
${\Lambda_m\over\Lambda_M}$ held fixed, simply choosing ${\tilde D} =4$ the
matching conditions
fix the renormalization conditions of the effective theory of the
light particle alone precisely, and we may choose any $D\geq 4$; this
generalizes the result of \lky\
to all $D>6$ with, it seems to us, remarkably little effort.
Conversely, if $\Lambda_0$ is not very much larger than $\Lambda_M$,
we must choose $D$ to be much the same as ${\tilde D}$, and
relatively little is gained in the precision of the light particle
theory by matching it to the heavy particle one. Of course, we expect
that in all realistic theories we will be somewhere between these two extremes.

Finally we consider stability.
We can use the technique described in \S 5 of \REFT\ to construct a
large class of nonperturbatively stable theories containing only
the light particle (with two scales $\Lambda_m\ll\Lambda_M$), and
also another large class containing both the light particle and a
stable heavy particle\foot{A single heavy particle cannot decay into
light ones because of the global $Z_2$ symmetry imposed on the
heavy particle field.} (now with three scales
$\Lambda_m\ll\Lambda_M\ll\Lambda_0$), each with the same
renormalization conditions on physically relevant couplings.
It should be clear that both these theories will have S--matrices
which, at least in perturbation theory, are both unitary and causal.
Furthermore the S--matrix of the latter is unitary not only on the
full space of light and heavy particle external states of arbitrary
energy, but also on the subspace
of light particle states with energies less than $2M$, since from
the cutting relation the heavy particles can only contribute to the
imaginary parts of light particle amplitudes above
threshold. There is no guarantee however that the light particle theories
on this subspace, when analytically continued to describe
light particle processes of arbitrarily high energies, will remain stable
non--perturbatively. However it is not difficult to see from \edxxxie\
that there always exists a stable light particle theory whose S--matrix is
arbitrarily close to that of the full theory when restricted to this
subspace.

\medskip

\nref\rvel{M.~Veltman, {\it Physica~}\vyp{29}{1963}{186}.}
We may now consider a wider class of scalar field theories.
In particular, if we were to relax the $Z_2$ symmetry of the heavy
particle interactions, they
would be free to decay into the light particles. We can only really do
this in the effective theory, since as explained in \S 6 of \REFT\ if
the vacuum is to be stable then the
couplings $g_M'$ and $g_{mM}'$ for vertices involving an odd number of
heavy particles are bounded above (by $(M/\Lambda_0)g_M^{1/2}$ and
$(M/\Lambda_0)g_{mM}^{1/2}$ respectively); in
the local limit the $Z_2$ symmetry is thus restored.
The unitarity and causality of theories in which a heavy
particle is allowed to decay into lighter particles was discussed
long ago by Veltman\rvel; the issue is a subtle one since in principle
only the light particles of such a theory may be considered as
external states, the heavy particle pole being no longer on the
physical sheet (though it approaches it in the limit $g_{mM}\to 0$).
However, it can be shown that there exists a unitary and causal S-matrix
for the scattering of the light particles alone.
This S--matrix restricted to the subspace of light particle
states with energies less than $M$ is manifestly unitary, and using
the methods of \REFT\ we may clearly construct an effective light
particle theory whose S--matrix tends arbitrarily close to that of
the full theory on
this subspace, and the improved decoupling theorem still
holds.

The question of stability is a little more complicated for
such theories. Veltman showed \rvel\ that the light field equations of motion
obtained by integrating out the unstable heavy particle give a
stable vacuum for the light field. However, the
manifestly stable light particle theory obtained from adiabatic
quantization \REFT\ may not be able to reproduce the light particle
theory obtained by formally integrating out the heavy field at
$\Lambda_M$, and then
expanding all nonanalytic terms in powers of momenta. This is because the
`higher derivative' terms generated via adiabatic quantization come
about purely from the radiative part of the renormalization group flow,
while when the heavy field is integrated out such operators are
generated from tree diagrams. Thus the `higher derivative' operators in the
latter may be larger than those we can feasibly produce in the former. This is
not really an important restriction for the present theory because the
coupling between the light and heavy particles is necessarily weak
compared to the couplings in the light particle theory due to the
stability condition.
However, in certain regions of coupling space, and for theories with
unstable particles in which there is no reason for the coupling
leading to the heavy particle decay to be relatively small (as for
example in electroweak theory) normal adiabatic quantization
may not be sufficient to
produce a suitable effective field theory. In these cases
it is however possible to obtain a manifestly stable primordial action by
first writing it in terms of both the heavy and light field and then
integrating out the unstable heavy particle. Expanding all nonanalytic
terms in powers of momenta, and matching their Taylor series (which
will all have radiuses of convergence of at least $M$) to analytic vertex
functions, we may then use adiabatic quantization to
produce a manifestly unitary and causal effective theory. It is not difficult
to see that at scales below $\Lambda_M$ this effective theory will be
identical (order by order in powers of momenta) to the manifestly
stable theory obtained by adiabatically quantizing the full
theory and then decoupling the heavy fields at $\Lambda_M$.

We can also consider a theory in which not only the $Z_2$ symmetry of
the heavy particle interactions is broken, but in which the $Z_2$
symmetry of the light particles is also broken at scales of order
$\Lambda_M$ by its interactions with the heavy particle. Thus while
the physically relevant renormalization conditions on Green's
functions involving only light particles are still set at
momenta of order $\Lambda_m$, they will no longer be $Z_2$ symmetric,
since amplitudes with an odd number of light particle legs may be
induced by corrections involving virtual heavy particles.
It is not difficult to see that the result \edxxxie, when generalized
to cover this softly broken case, now implies that the
amputated connected Green's functions of the light particle theory
with an odd number of legs $2m+1$ must be suppressed relative to those
with $2m$ legs by a factor of $(\smallfrac{\Lambda_m}{\Lambda_M})$.
This result is a simple paradigm for the breaking of parity by the
weak interaction.

\newsec{The Local Limit.}

\nref\riii{S.~Weinberg, {\it Physica~}\vyp{96A}{1979}{327}\semi
J.~Gasser and H.~Leutwyler, \AP\vyp{158}{1984}{142};
\NP\vyp{B250}{1985}{465}.}
When working with an effective quantum field theory it is usually simplest
both conceptually and technically to keep the regularization scale
$\Lambda$ below the naturalness scale $\Lambda_0$; indeed when
constructing stable theories as described in \S 5.2 the theory was
only formally quantized for $\Lambda\lsim\Lambda_0$. This was
particularly obvious in our discussion of decoupling in the previous
section, where the naturalness scale $\Lambda_M$ of the light particle
theory was identified with the mass of some new heavy particle. When
formulated in this way, we are always free to choose the number of physically
relevant couplings we wish to work with, quite independently of
technical details such as the order at which we truncate our
perturbation theory.

Of course it is not necessary to follow this approach; as Green's
functions are by construction independent of $\Lambda$ we may use
the full renormalization group equations to take
$\Lambda/\Lambda_0$ to infinity with impunity. Furthermore as explained in
\REFT\ all physical quantities are independent of the from of the
regulating function $K_\Lambda$, so in this limit it might seem
reasonable to introduce an alternative regularization procedure, for
example dimensional regularization. The effective theory would then
closely resemble a local quantum field theory, and indeed this is the
approach which is most frequently to be found in the original
literature\refs{\eftdec,\riii}.
We feel that this approach obscures the relationship between
physics at different scales and in particular more and more
counterterms must be introduced as we work to higher and higher
loop order, simply in order to render the theory finite (thus
destroying some of the properties of conventional local field
theory, such as manifest stability). Indeed none of
the conjectures formulated in the framework of local effective
theories have, to our knowledge, been rigorously proven.

We are perhaps best able to see how our approach relates to this
`conventional' approach if we consider it in the context of
decoupling. In \S 2 we can consider the theory containing the two
particles in the local limit $\Lambda_0/\Lambda_m\to\infty$; this does not
change \S 3 at all if we
consider the effective theory containing only the light particle to
have naturalness scale $\Lambda_M$. More relevantly, we could
instead set the boundary conditions on the irrelevant couplings of the
light particle theory at $\Lambda_0$, and thus
also take this theory in the local limit; the bounds lemma 1 will be
unchanged because the low energy renormalization conditions
on the relevant couplings are independent of $\Lambda_0$, and the
proof of decoupling, lemma 11, goes ahead just as it did in \S 3.

The case for the improved decoupling theorem presented in \S 8.3
is not so simple. In order to match the physical renormalization
conditions on the vertices with dimension $D>4$ in the light
particle theory to those of the theory containing the heavy particle,
they must be such as to be natural when evolved back to $\Lambda_M$.
This means that they cannot remain natural if we evolve them above
$\Lambda_M$ to $\Lambda_0$, but rather will begin to diverge as powers
of $\Lambda_0/\Lambda_M$ (up to logarithms). In the local limit,
we will thus find a
theory with infinite counterterms which are
just such as to guarantee that the renormalized insertions in the
light particle theory reproduce the effects of the heavy particle.
The light particle theory is then an effective theory in the
conventional (local) formulation.

Unfortunately, the combination of scales (or scales and factors of
$\epsilon^{-1}$ if one were to use dimensional
regularization) now makes the scale dependence of the physics of the
effective theory rather more difficult to disentangle. In particular,
it is in general necessary to work with the full infinite set of
counterterms, even though after renormalization operators with
dimension greater than $D$ are discarded; even if the renormalized
couplings corresponding to these operators were set to zero, the form
of the flow equations means that they
will be nonzero when evolved back to $\Lambda_0$, and infinite in the
local limit.

Consider, for example, computations in the loop expansion. The form
of the flow equations shows us that to create the natural
effective Lagrangian at $\Lambda_M$ from that at $\Lambda_0$ we
can only decrease the number of legs of a vertex by forming
loop diagrams, and each loop causes a maximum reduction of two legs.
Therefore, a vertex with $2m +2n$ legs at $\Lambda_0$ will make a
contribution to a vertex at $\Lambda_M$ at a power of
$\hbar$ which is at least $n$ higher. So even if we only wish to work to
an accuracy of order $({\Lambda_m\over\Lambda_M})^{D-2}$, and
therefore at retain renormalized vertices with at most
$D$ legs, if we work at order $l$ in the loop expansion it will
be necessary to include vertices at $\Lambda_0$
with $2l+4$ legs, just to render the renormalized vertices finite. At
arbitrarily high order, we thus need an arbitrarily large number of
counterterms. Outside perturbation theory the number of counterterms
would in general be infinite.

Of course, there may well be situations where for practical reasons
the local formulation is advantageous. For example when making
perturbative computations in
theories with local or nonlinearly realized symmetries (as in \riii)
dimensional regularization is convenient because it manifestly
preserves such symmetries; in the flow equations they seem to be
necessarily broken by the introduction of the regularization
function.\foot{Though clearly some complicated remnant of the symmetry
must remain.} Similarly, nonperturbative computations in an effective theory
could perhaps be performed on the lattice if the lattice spacing
$a\ll\Lambda_0\inv$ (though there would in general be problems with
spurious instabilities on the scale of the lattice spacing if the
finite difference equations were truncated at finite order). But
formal issues, which in general benefit from
a clear separation of scales are, we feel, most readily
formulated and proven within the quasi--local formulation of effective
field theory as presented in \REFT.

\newsec{Predictivity.}

To summarize, we have shown that a theory of two scalar particles with
$Z_2\times Z_2$ symmetry and well separated mass scales
$\Lambda_m\ll\Lambda_M\ll\Lambda_0$ may be approximated arbitrarily
closely at scales below $\Lambda_M$ by an effective theory containing the
light particle alone. This decoupling theorem, which we proved to all
orders in perturbation theory, may be readily extended to any theory
of scalar and/or spin half particles, with global symmetries, since as
explained in \S 6 of \REFT\ no new ideas are then necessary. All that
is required is that the particles are arranged in a hierarchy of mass
scales $\Lambda_m\ll\Lambda_{M_1}\ll\Lambda_{M_2}\ll\cdots\ll\Lambda_0$; each
scale may then be decoupled sequentially. The decoupling theorem will
also remain valid for effective field theories set in space--times
with dimensions other than four, such as Kaluza--Klein
theories. Theories with spontaneously
broken symmetries, nonlinearly realized symmetries, or local
symmetries are more difficult to handle, however, and indeed for them the
decoupling theorem may sometimes break down.

We consider finally the opposite problem of attempting to predict the
existence of new particles at higher mass scales than current
experiments. We showed in \REFT\ that we can describe the physics of
processes below a certain energy scale $\Lambda_M$, using an effective
field theory containing only those fields which correspond to
particles with masses $m$ below this scale. As we
increase the energy at which the physics is probed, we need more
parameters to maintain the same precision. If we were to make no
assumptions about the nature of the theory at $\Lambda_M$, all of these
new parameters would be constrained only by naturalness. However, if
there were a new particle at $\Lambda_M$, the new parameters would be
more tightly constrained by the requirement that at $\Lambda_M$ the
light particle theory matched to the new theory containing the extra
particle. The higher the new naturalness scale, $\Lambda_0$, the
tighter these constraints would be, and the easier it would be to guess
the existence of the heavy particle from the accurate examination of
light particle processes below its threshold; if $\Lambda_0$ were too
low, it would be difficult to disentangle the effects at low energy of
the heavy particle from those of the physics at $\Lambda_0$. Of course
the new couplings will be easier to measure accurately if they break
some global symmetry (as for example the $Z_2$ symmetry considered
above); the observation of parity violating weak interactions, through
both weak and neutral current interactions, was seen as good evidence
for the existence of intermediate vector bosons, and likewise
new insights are expected from the study of CP violation (although here
there is as yet no compelling candidate for an underlying theory).

If the new heavy particle really existed, then as we
reached the threshold for heavy particle production, our effective theory
containing just the light particle would necessarily breakdown, since
by construction it does not contain the threshold singularity.
Although by using enough parameters, the light particle effective
theory can mimic a
more fundamental theory to any order in the Taylor expansion of its
light particle amplitudes, when there is a genuine new particle of
mass $M$ in the more fundamental theory this Taylor expansion has
radius of convergence $M$. The light
particle theory would then have to be discarded, at least for processes at and
above threshold.

However it could in principle turn out, despite previous circumstantial
evidence, that when the scale $\Lambda_M$ is reached there is no new
particle at all, and the light particle theory can
still be used there (albeit with limited predictivity)
\foot{It could be, for example, that the light particles are
composite, as is the case is chiral perturbation theory. In this case
too it would, at least in principle, be possible to compute the
infinite number of coupling constants in the light particle effective
theory in terms of those of some new theory of the constituents.}; it is
only possible, given the fact that all the low energy
experimental data are in practice limited both in accuracy and
quantity, to be certain of the existence of a new particle
by actually producing it. Seeing is believing.

\bigskip
\noindent{\bf Note Added.}\nobreak
\medskip\nobreak
Recently two preprints have appeared \ref\gzk{L.~Girardello and
A.~Zaffaroni, IFUM/432/FT, SISSA/14/94/EP\semi
C.~Kim, SNUTP 94-20.} which also attempt to prove the decoupling
theorem using the exact renormalization group. Both these authors work
implicitly in the local limit $\Lambda_0\to\infty$, however, and
unfortunately set artificial renormalization conditions on the heavy particle
vertices at zero momentum, which renders the proof of decoupling relatively
trivial, at the expense of substantially reducing its content.

\bigskip\vfill\eject
\noindent{\bf Acknowledgements.}\nobreak
\medskip\nobreak
We would like to thank the Royal Society and the SERC for financial
support during the period when most of this work was done, and the
Theoretical Physics group of Oxford University for their hospitality;
RST would also like to thank Jesus College, Oxford and the
Leathersellers' Company for a graduate scholarship.

\footatend\vfill\supereject\immediate\closeout\rfile\writestoppt
\baselineskip=14pt\centerline{{\bf References}}\bigskip{\frenchspacing%
\parindent=20pt\escapechar=` \input refs.tmp\vfill\eject}\nonfrenchspacing\vfill\eject
\end